\PassOptionsToPackage{authoryear}{natbib}
\documentclass[fleqn,usenatbib]{mnras}

\usepackage{multirow}
\usepackage[T1]{fontenc}
\usepackage{ae,aecompl}
\usepackage{amsmath}
\usepackage{etoolbox}
\usepackage{breqn}
\usepackage{graphicx}
\usepackage{textcomp}
\usepackage{subfigure}
\usepackage{upgreek}
\usepackage{pgffor}
\usepackage{color}
\usepackage{nicefrac}
\usepackage[normalem]{ulem}
\usepackage{diagbox}

\usepackage{amssymb}    
\usepackage{tablefootnote}

\PassOptionsToPackage{authoryear}{natbib}
\usepackage{natbib}
\bibpunct{(}{)}{;}{a}{}{,}
\PassOptionsToPackage{english}{babel}
\usepackage{babel}

\renewcommand{\mu}{\upmu}

\newcommand{\ssc}{_\textsc{\tiny ssc}}
\newcommand{\syn}{_\textsc{\tiny syn}}

\newcommand{\ma}{_\textsc{\tiny max}}

\makeatletter
\def\dif{\@ifnextchar[{\@with}{\@without}}
\def\@with[#1]#2{
  \ensuremath{\frac{\foreach \x in {#2}{\mathrm{d}\x\,}}{\foreach \x in {#1}{\mathrm{d}\x\,}}}
}
\def\@without#1{
  \ensuremath{%
    \ifx\hfuzz#1\hfuzz
    \mathrm{d}
    \else
    \foreach \x in {#1}{\mathrm{d}\x\,}
    \fi
    }
}
\def\mitya{\@ifnextchar[{\@mwith}{\@mwithout}}
\def\@mwith[#1]#2{
  {\color{blue} \bfseries #2}\footnote{Comment: #1}
}
\def\@mwithout#1{
  {\color{blue} \bfseries #1}
}
\makeatother

\PassOptionsToPackage{hyphens}{url} 
\title[The Crab nebula variability with CTA]{The Crab nebula variability at short timescales with the Cherenkov Telescope Array }
\author[E. Mestre et al.]{E.~Mestre$^{1,2}$,
E.~de O\~na Wilhelmi$^{1,2,3}$,
D.~Khangulyan$^{4}$,
R.~Zanin$^{5}$,
F.~Acero$^{6}$,
\and D. F.~Torres$^{1,2,7}$
\\
$^{1}$Institute of Space Sciences (ICE/CSIC), Campus UAB, Carrer de Can Magrans s/n, 08193 Barcelona, Spain\\
$^{2}$Institut d' Estudis Espacials de Catalunya (IEEC), 08034 Barcelona, Spain\\
$^{3}$Deutsches Elektronen Synchrotron DESY, 15738 Zeuthen, Germany\\
$^{4}$ Department of Physics, Rikkyo University, Nishi-Ikebukuro 3-34-1, Toshima-ku, Tokyo 171-8501, Japan\\
$^{5}$CTA Observatory GmbH, Via Piero Gobetti 93, I-40129 Bologna, Italy\\
$^{6}$ AIM, CEA, CNRS, Universit\'e Paris-Saclay, F-91191 Gif-sur-Yvette Cedex, France \\
$^{7}$Instituci\'o Catalana de Recerca i Estudis Avan\c{c}ats (ICREA), E-08010 Barcelona, Spain
}

\begin{document}
\maketitle
\date{Received  ; accepted }

\pubyear{2018}


\begin{abstract}
Since 2009, several rapid and bright flares have been observed at high energies (>100 MeV) from the direction of the Crab Nebula. Several hypotheses have been put forward to explain this phenomenon, but the origin is still unclear. The detection of counterparts at higher energies with the next generation of Cherenkov telescopes will be determinant to constrain the underlying emission mechanisms. We aim at studying the capability of the Cherenkov Telescope Array (CTA) to explore the physics behind the flares, by performing simulations of the Crab Nebula spectral energy distribution, both in flaring and steady state, for different parameters related to the physical conditions in the nebula. In particular, we explore the data recorded by \emph{Fermi} during two particular flares that occurred in 2011 and 2013. The expected GeV and TeV gamma-ray emission is derived using different radiation models. The resulting emission is convoluted with the CTA response and tested for detection, obtaining an exclusion region for the space of parameters that rule the different flare emission models. Our simulations show different scenarios that may be favourable for achieving the detection of the flares in Crab with CTA, in different regimes of energy. In particular, we find that observations with low sub-100\,GeV energy threshold telescopes could provide the most model-constraining results.   
\end{abstract}

\maketitle

\section{Introduction}

The Crab nebula has been used as standard candle in gamma-ray astronomy since the first Imaging Air Cherenkov Telescopes (IACTs) began to operate. The science verification period of the current and future IACTs rely on observations of the, in principle, stable gamma-ray flux detected in the bright nebula, and it has been used to characterise the performance of different instruments (\citealt{2000Aharonian,Aleksic2015,2015Holler,2015Meagher,2016Aleksic,2019Abeysekara,2019Amenomori}). However, several flares of different magnitude have been detected in the last years with spaceborne gamma-ray instruments (\citealt{Abdo2011,Tavani2011,Buehler2012,2012Ojha,2013Mayer,2013Striani,2020arXiv200507958A}) in the high-energy regime (HE, \(\geq100\rm\,MeV\))
with variability time scales of hours (\citealt{Abdo2011}, \citealt{Tavani2011}, \citealt{Balbo_2011}). 
During these flaring periods, the flux of the nebula shows rapid variations, releasing a huge amount of energy: for example, during the April 2011 flare, first detected by \emph{Fermi}-LAT \citep{2011ATelB,Buehler2012} and later confirmed by AGILE \citep{Striani2011}, the nebula doubled its HE flux level with respect to the steady state in less than 8 hours. A review of the different flares detected up to September 2013 can be found in \cite{2014Buhler}. The rapid variability and the energy range in which the flares are detected point to a phenomenon associated to fast variation of magnetic fields and/or compact regions. These two ingredients do not favour a rapid variation (at a detectable level with the current generation of Cherenkov telescopes) in the inverse Compton (IC) component that emerges in the TeV regime \citep{2004Horns}. A large effort was done to follow those flares in several multi-wavelength campaigns \citep{Weisskopf2013,2014HESS} to search for a new ingredient that could shed light on the mechanisms underlying the flares observed. These observations did not result in any positive correlation between spectral and/or morphological variations of the nebula and the (hundreds of) MeV flares.

Several theoretical works have been put forward to explain the characteristics of the observed emission, which results in different predictions of the flux level at different wavelengths. In particular, the GeV regime can be accessed by current and future IACTs with great sensitivity \citep[CTA,][]{2013CTA,2019Natur.575..455M,2019Natur.575..464A}. 

The detection of transients is one of the Key Science Programs (KSPs) of the Cherenkov Telescope Array \citep[CTA,][]{2013CTA}. It comprises the ability to rapidly respond to a broad range of multi-wavelength alerts from other observatories, being the design of the telescopes optimized for rapid movement.
In this paper, we explore the capability of CTA to constrain the flare contribution in the GeV and TeV regime. For that purpose, we performed a number of simulations to reproduce the $\sim$400\,MeV emission detected by gamma-ray satellites, and derived the expected emission in the high energy regime, under different conditions of magnetic and photon fields, and particle spectral energy distribution. The comparison with the CTA simulated data \citep[see][]{2019Mestre}, and previous IACTs observations \citep{2014HESS} result in constraints on the physical parameters ruling the flare emission.

The paper is structured as follows: in Section \ref{sec2} we describe the analysis technique, including the parameters and variables chosen for the simulations. We show the results of the simulations in Section \ref{sec3}. In Section \ref{sec4}, we discuss the potential of CTA to constrain different theoretical scenarios proposed to explain the flare emission. Finally, in Section \ref{sec5} we emphasise some of the conclusions we reached in this work.

\section{Simulations and Methods}
\label{sec2}

To evaluate the capability of CTA to explore the properties of the population of electrons behind the flares in the Crab nebula, we first explore its basic physical parameters (in Section \ref{sec2.1}) and the properties of its flares (Section \ref{sec2.2}). Second, we fit the flare spectrum assuming a classical synchrotron emission as coded in the \textsc{naima} \textsc{python} package \citep{naima} and calculate the corresponding IC component for different realizations of magnetic field strengths (Section \ref{sec2.3}). Finally we convolve the synchrotron tail and IC components with the CTA response, including the expected emission from the nebula and derive the CTA sensitivity for different population of electrons compatible with the flare emission observed (Section \ref{sec2.4}). 

\subsection{General physical properties of the Crab nebula}
\label{sec2.1}

The broad-band spectrum of the nebula consists of a wide synchrotron component and a narrow IC peak. Synchrotron radiation is expected to be the dominant channel in which particles cool down. Therefore, the radiation observed critically depends on the strength of the magnetic field. The average magnetic field strength in the nebula is well-constrained through the multi-wavelength properties of the Crab Nebula to be $\bar{B} \sim 120 \upmu {\rm G}$ \citep[see,e.g.,][]{1996Atoyan,1998ApJ...503..744H,2012Martin,2020MNRAS.491.3217K}. The flares can occur however in particular locations of the nebula, where the magnetic field can be very different from the average one \citep[see, e.g.,][]{1984ApJ...283..710K,2012MNRAS.427.1497L,2014IJMPS..2860168P}. 
The flare spectral index also shows a large variability from flare to flare \citep{2014Buhler,2020Arakawa}, which results on a large range of particle indices to consider, when calculating the parent particle population. 

The IC component emerges from high energy electrons up-scattering several photon targets: the cosmic microwave background (CMB), far-infrared (FIR), near-infrared (NIR) photon background fields, and the synchrotron emission, which is believed to dominate the total gamma-ray emission of the nebula. The FIR background is typically attributed to an isotropic photon field emitted by dust, at a temperature of 70 K with an energy density of 0.5 $\rm{eV}/\rm{cm}^{3}$, while the NIR background photon field is usually described for starlight with temperature of 5000 K and energy density of 1 $\rm{eV}/\rm{cm}^{3}$ \citep{1996Atoyan}. These background fields can be approximated as diluted Planckian distributions and the corresponding emission spectra can be obtained based on an approximate treatment \citep{2014ApJ...783..100K}. The Synchrotron Self-Compton (SSC) component however depends on the size of the emission region. For the total nebula, with a radius of $R\syn\sim1.5\rm\,pc$, the total energy density provided by the synchrotron photons amounts:

\begin{equation*}
\centering
\omega\ssc = \frac{2.24L\syn}{4\upi c R^{2}_{syn}} \simeq 2\ {\rm eV}\ {\rm cm}^{-3}
\end{equation*}
(where the factor 2.24 is obtained for a homogeneous spherical source, \citealt{1996Atoyan}), for a total luminosity in the synchrotron regime of the nebula of $L\syn\sim 10^{37}$erg\,s$^{-1}$ (see Table 1 in \citealt{1997Aharonian}). The fast variability observed on the flares nonetheless, limits the size of the emission regions (due to causality arguments). For a duration of $t_{var}$, the size of the emission region should be limited to $c\times t_{var}$.  



Considering the uncertainties described above, we simulate our particle population using a range of particle indices between $\Gamma_{\rm{e}}$=[1--3] (associated to non-thermal acceleration processes, \citealt{Longairbook}). To calculate the corresponding synchrotron emission we use a relatively broad range of magnetic field intensities, for particular regions in which the flares might occur, spanning from a few $\rm \upmu G$ and $\rm mG$. Likewise, the IC part is calculated using the CMB and IR photon targets, whereas the SSC contribution is obtained for each particular flare duration considered. Finally, the space of parameters is also a priori limited by the total energy budget stored in the nebula. The latest can be estimated as the product of the nebula luminosity in $\gamma$-rays ($L_{\gamma} \sim 2\times 10^{35}\ \rm{erg}\,\rm{s}^{-1}$, see \citealt{1998Rudak}) and synchrotron cooling time ($\tau\syn$) for the electrons (with energy $E_{e}$) of the nebula: 
\begin{equation}
\label{eq:recon_Lorentz}
\tau\syn L_{\gamma} = \frac{3m_{e}^{4}c^{7}}{2e^{4}} E_{\rm{e}}^{-1} B^{-2} \sim 5\times 10^{43} (B / 100\,\mu{\rm  G})^{-2}~\rm{erg}
\end{equation}
Note however, that this energy limit does not account for re-acceleration of particles, that can result on additional boosts of energy.

 \begin{figure*}
  \centering
  \includegraphics[width=0.8\textwidth]{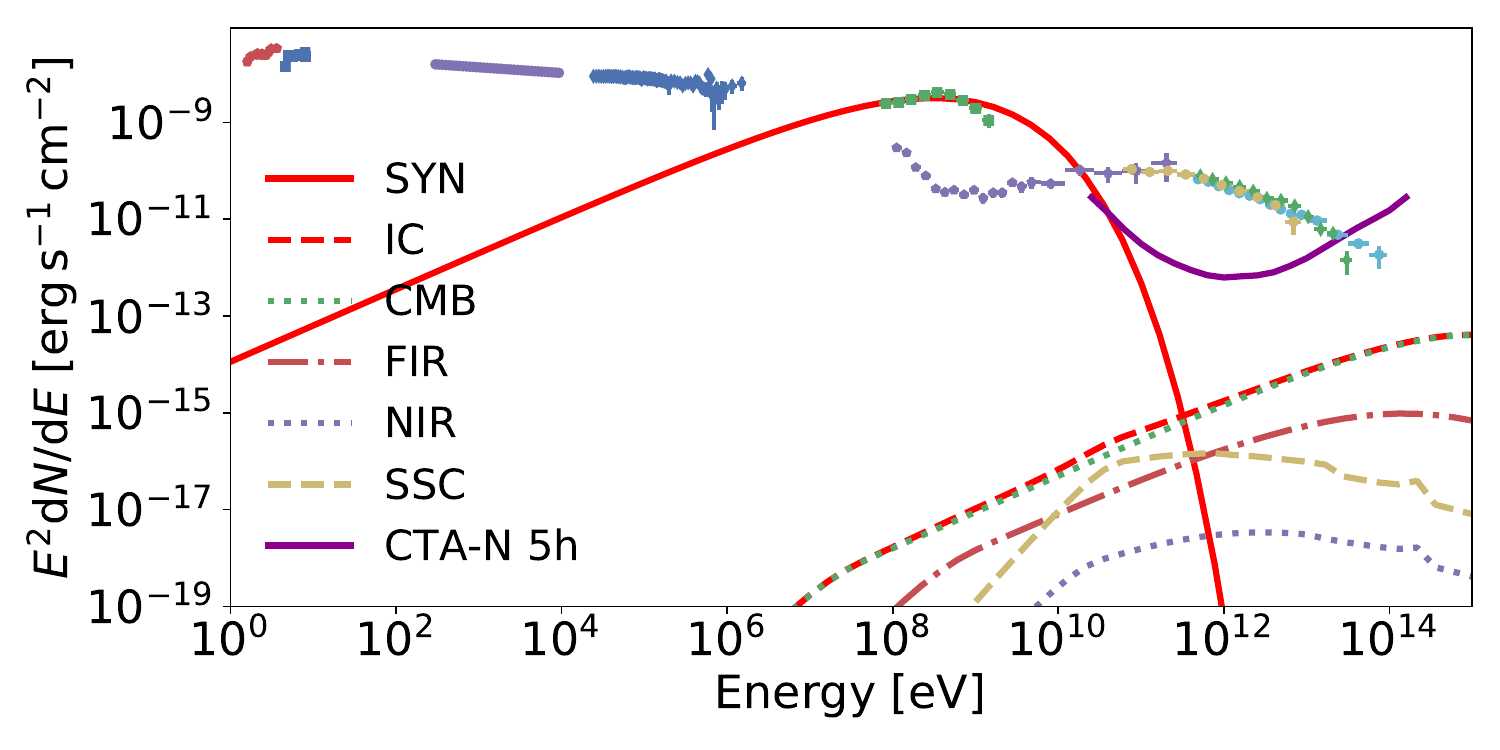}\\
  \caption{Synchrotron and IC of a simulated flare (fitted to the 2011 April flare spectrum) with $\rm{B}= 100$ $\upmu$G and $\Gamma_{\rm e} = 1.5$. The synchrotron and total IC emissions are plotted in red solid and dashed line, respectively. The rest of the lines represent the IC with CMB, FIR and NIR background photon fields and the SSC. The magenta solid line represents the sensitivity of the CTA northern array (for 5\,h of observation time, see the instrument response functions\protect\footnote[1]{}). The green squares correspond to the nebula spectrum during the 2011 April flare (as seen by \emph{Fermi}). The rest of the data correspond to the steady nebula emission (the compilation was taken from \citealt{2010A&A...523A...2M} and \citealt{Buehler2012}).}
  \label{fig:Flareexample}
\end{figure*}

\subsection{Gamma-ray properties of the Crab Nebula flares}
\label{sec2.2}

We performed the simulations for two flares with different characteristics: a very bright one with parameters similar to the one observed by \emph{Fermi}-LAT in April 2011 \citep{Striani2011,Buehler2012}, and a moderated-flux flare as the one observed in March 2013 \citep{2012Ojha, 2013Mayer}. For the latest, simultaneous observations with the H.E.S.S. and VERITAS experiments were performed \citep{2014HESS,2014Aliu}, which allow us to test the limits on the parameters space with the current IACT sensitivity. 
In the following we briefly describe the main characteristics of both flares. The data points are extracted from \citealt{2014Buhler}. 

The gamma-ray flare detected in April 2011 \citep{Striani2011,Buehler2012} lasted nine days, reaching a peak photon flux of $(186 \pm 6) \times 10^{-7}\ \rm{cm}^{-2}\rm{s}^{-1}$ above 100\,MeV, implying a flux enhancement by approximately a factor 30 compared to the average flux from the Crab nebula \citep{Buehler2012}. The $\gamma$-ray flares from the Crab show complex substructures with sub-flares of duration of a few hours. For instance, the April 2011 flare had two sub-peaks centered around the dates 55665 and 55667 (in MJD), both with a doubling timescale ($t_{var}$) smaller than 8\,h, implying (because of causality arguments) a compact emission region smaller than $c \times t_{var} \sim 2.8 \times 10^{-4}$ pc in size. The flare spectral energy distribution (SED) had a distinctive narrow shape, peaking at $E_{peak}\simeq 400$ MeV. 

The Crab flare of March 4, 2013 was reported by \emph{Fermi}-LAT \citep{2012Ojha, 2013Mayer} and AGILE \citep{2013Striani} when the peak photon flux of the synchrotron emission for energies above 100\,MeV was $(103.4\pm 0.8) \times 10^{-7}\ \rm{cm}^{-2}\rm{s}^{-1}$, about 17 times above the level of the steady component. The flare lasted for approximately two weeks and the variability was measured on timescales of $\sim 5$ hours, which results on a region size of $\lesssim1.7 \times 10^{-4}$ pc \citep{2013Mayer}. Interestingly, the SED peaked at an energy of $\sim 400$\,MeV at the time of the highest flux, as in the case of the April 2011 flare.
Observations of the flare were carried out with different instruments, providing the opportunity to study the emission during the flaring state at multiple wavelengths, from infrared to X-rays \citep{2013Mayer} and also in the very high energy (VHE) regime \citep{2014Aliu,2014HESS}. A blind search of flares in Crab by ARGO-YBJ air shower detector reported no significant excess of events during the 2011 and 2013 flares \citep{2015Bartoli}.
The observations conducted with VERITAS and H.E.S.S. the following days after the 2013 flare reported also no significant changes in the flux of the nebula (above 1\,TeV). Considering systematical and statistical errors, both observations result in similar upper limits to the variability of the integral flux of $\sim 55-65\%$ (for a 95\% CL). We used a fiducial value of 60\% to compare with our simulations. 


\subsection{Simulation of the electron particle distribution}
\label{sec2.3}

The electron particle distribution simulated to reproduce the observed flares is characterized by a power-law distribution with an exponential cutoff:
\begin{equation}
\frac{dN_{e}}{dE} = N_{0}\left(\frac{E}{1 {\rm TeV}}\right)^{-\Gamma_{\rm e}}\exp\left(-\frac{E}{E_{p}}\right) 
\label{eq:ecplformula}
\end{equation}
The magnetic field (B) and the particle index ($\Gamma_{\rm e}$) are the only parameters left free in order to derive the resulting SED of synchrotron radiation from the electron population. The maximum energy reached, the amplitude of the gamma-ray spectrum (which is determined by the amplitude of the electron particle distribution, $N_{0}$), and the particle spectrum cutoff energy are obtained from fitting the spectrum at hundreds of MeV to the different data sets employed (the LAT measurements of the April 2011 and March 2013 flares). The fitting is performed by means of a log-likelihood optimization method, implemented using the open-source software \textsc{naima} \citep{naima}.

We compute, first, the amplitude ($\rm{N}_{\rm{0}}$) and the cutoff energy of the particle spectrum ($E_{p}$), for the chosen particle index and magnetic field, by fitting the resulting synchrotron emission to the Fermi-LAT flare data.

We considered $\Gamma_{\rm e}$ ranging from 1 to 3 as argumented above (in linearly spaced bins of 0.1) and the magnetic field (B) ranging from 10 $\mu$G to 5\,mG in 14 bins (10 $\mu$G, 100 $\mu$G and 1\,mG, plus 11 values logarithmically spaced from 50 $\mu$G to 5\,mG). Note that both very low ($10 \mu $G) and very large (few mG) magnetic fields are difficult to justify in standard pulsar wind nebula (PWN) theory. However, we include them to cover all possible hypotheses, and to probe the performance of the simulations over a very large parameters space. In the discussion section, we will only focus then on results for $B > 50$ $\mu$G.

\begin{figure}
  \centering
  \includegraphics[width=0.5\textwidth]{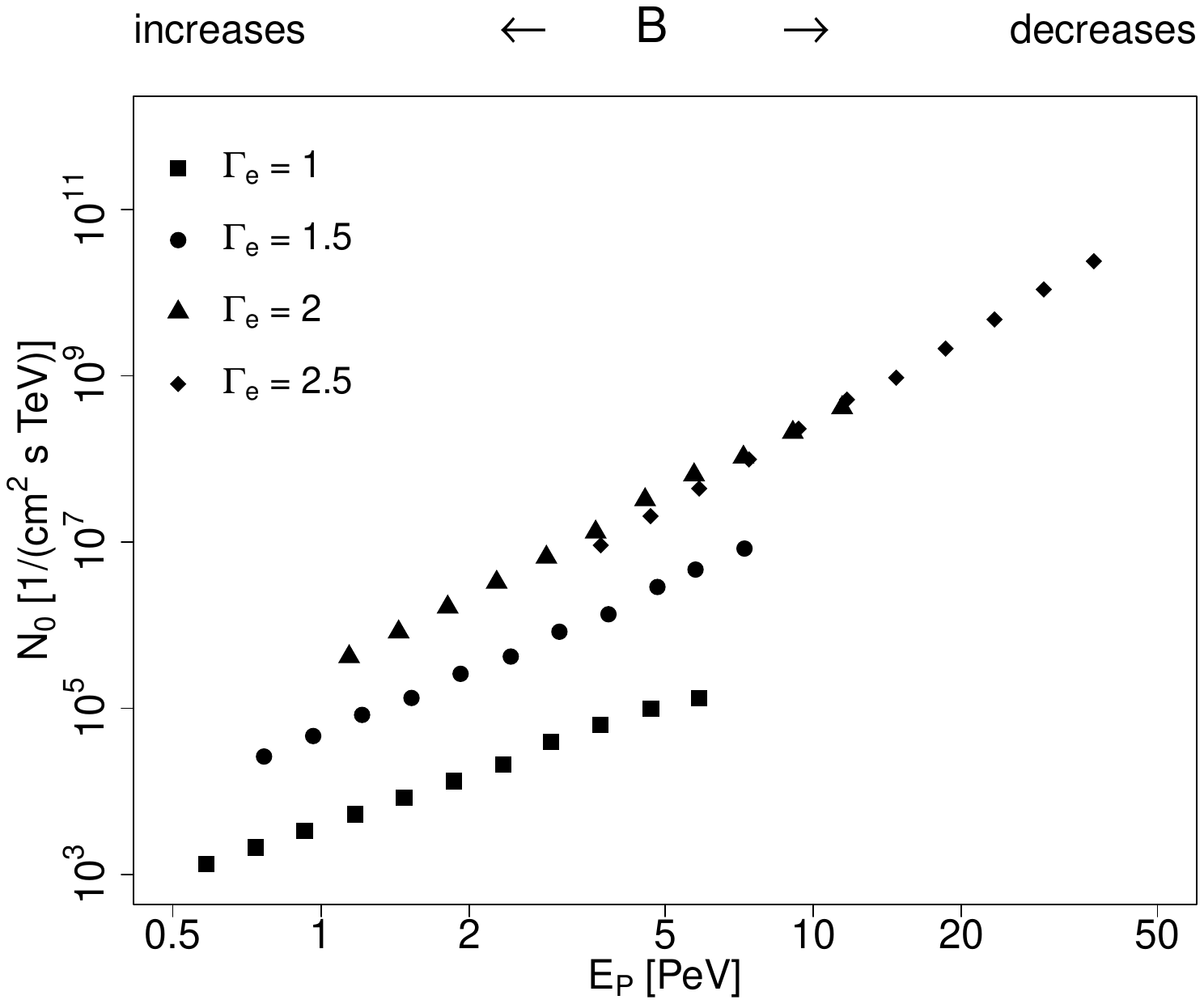}\\
  \caption{The fitted amplitude of the electron energy distribution ($\rm{N}_{\rm{e}}$), with respect to $\Gamma_{\rm e}$, $E_{p}$, and B for the April 2011 flare. For each value of $\Gamma_{\rm e}$, the points correspond from left to right to 11 values of magnetic field logarithmically spaced from 5\,mG to 50 $\upmu$G.}
  \label{fig:Particlespectrumamplitude}
\end{figure}

The synchrotron emission is computed using the \textsc{naima} \textsc{python} package \citep{naima}, according to the approximation of the synchrotron emissivity in a random magnetic field in \citealt{2010PhRvD..82d3002A}. We selected a minimum electron energy of 50\,GeV. The amplitude and cutoff energy of the resulting emission spectrum for each fixed value of $\Gamma_{\rm e}$ and B is fitted to the 2011 (or 2013) flare observations. Using the resulting amplitude, we obtain the total density of electrons $\rm{N}_{\rm{0}}$ and compute the corresponding IC on all relevant photon targets.
The total SED of the flare (see Figure \ref{fig:Flareexample} as an example) is obtained as the sum of these different contributions. Both the synchrotron and IC are computed for 100 bins of energy logarithmically spaced from $10^{-7}$ eV to 1\,PeV. Finally, from the IC component integration, we obtained the total energy in electrons above 1\,TeV ($W_{\rm{e}}$) for the different flare models, which can be directly compared with the upper limit estimated in Section \ref{sec2.1}.

To illustrate the effect of the different parameters on the simulation, we plot in Fig. \ref{fig:Particlespectrumamplitude} the obtained amplitude ($\rm{N_{0}}$) with respect to the fitted cutoff energy ($\rm{E}_{\rm{p}}$), for different values of $\Gamma_{\rm e}$.

Additionally, we considered an alternative approximation for which we used a power-law decay for the synchrotron radiation model, instead of an exponential one \citep[as expected for jitter radiaton,][]{Kelner2013}. In particular, we fitted the \emph{Fermi} data above 400\,MeV with simple power-law models of fixed index ($\Gamma_{\rm{Jit}}$), ranging from 2.5 to 3.5 in bins of 0.1, and fitted the amplitude of the emission normalized to 0.85\,GeV of energy.

\subsection{Simulation of the CTA observations}
\label{sec2.4}

The simulations of the observations were performed folding the result from the radiation model with the CTA Instrument Response Functions (IRFs, \citealt{2015Hassan}, version prod3b\footnote[1]{The version prod3b of the IRFs and the CTA science performance requirements are public and available in https://www.cta-observatory.org/science/cta-performance/}). The later was done using the CTA \textsc{gammapy} \textsc{python} tool (version 0.7, \citealt{2017gammapy}). We used the IRFs corresponding to observations at low zenith angle (20\degr{}) from the northern array (aiming to guarantee the lowest energy threshold possible), and optimized for an exposure time of 5\,h. The results of the simulations are presented here for a fiducial observation time of 10\,h. 

The resulting spectrum is then reconstructed by means of a fitting process, using maximum likelihood (implemented in \textsc{sherpa}, see \citealt{1979Cash}) and Nelder-Mead Simplex optimization method \citep{NEADMED2,NEDMEAD1} based on a forward-folding technique (\citealt{2001Piron}). 
To avoid possible problems arising from the discretization of the radiation model, we parametrized the result of our model using an exponential power-law function (from 20\,GeV to 200\,GeV) and a power-law one (from 1.25\,TeV to 50\,TeV) to account for the MeV/GeV synchrotron contribution and the IC one at TeV's, respectively. The detection of flares at energies from 200\,GeV to 1\,TeV is considerably more difficult to achieve, since both the synchrotron and IC emissions from the flare are expected to be several orders of magnitude dimmer than at tens of GeVs and TeVs, respectively (see Fig. \ref{fig:Flareexample}).

In our simulations, the background is dominated by the emission from the nebula, specially in the TeV regime. To account for that, we used the results from the simulations presented in our previous work \citep[][]{2019Mestre}. In particular, we used the spectral shape described as a log-parabola as in \citealt{Aleksic2015}. The cosmic-ray background is provided by the CTA IRFs and computed in the simulations of the nebula, both in flaring and steady state. To evaluate the capability of CTA to disentangle variations on the large Crab nebula steady emission, we proceed as following: 
\begin{itemize}
    \item We fit the simulated spectrum (flare plus steady emission) in the GeV (i.e from 20\,GeV to 200\,GeV) and TeV regimes (from 1.25\,TeV to 50\,TeV) to simple power-laws.
    \item We compute the integral flux in both GeV and TeV regimes, using the best-fitted models.
    \item We compute the mean total expected excess (i.e., expected counts from the source, with background subtraction) and its standard deviation, by iterating over $10^{4}$ realisations for each model of flare (and 10\,h of observation time).
    \item The integral flux and mean expected excess in both GeV and TeV regimes are also computed for the flare and steady components individually, to use them as reference.
    \item To compare the non-flaring and flaring SED, we use a Pearson's chi-squared test, being the excess distribution the observed data ($\rm{H}_{1}$), to be compared to the steady state as expected data (null hypothesis $\rm{H}_{0}$). This test is then done for each of the $10^{4}$ realisations. We consider that a flare is detected when the null hypothesis can be rejected at $99$\% CL. 
    \item Finally, we compute the enhancement of integral flux above 1\,TeV (with respect to the simulations of the nebula in steady state) expected for the different models of flares, defined as:

\begin{equation}
\mathrm{Z} = \frac{\int_{1\rm TeV}^{300\rm TeV} \mathtt{F_{flare,E}}\,\mathrm{d}E + \int_{1\rm TeV}^{300\rm TeV} \mathtt{F_{Steady,E}}\,\mathrm{d}E}{\int_{1\rm TeV}^{300\rm TeV} \mathtt{F_{Steady,E}}\,\mathrm{d}E}\,.
\label{eq:zparameter}
\end{equation}

Note that models representing flares with very different spectral shapes can result in similar flux enhancements in Eq. \ref{eq:zparameter}.
 \end{itemize}

Furthermore, to test the capability of CTA to detect flares of different duration from the ones described above, we compute the minimum flare duration (using as fiducial flux level the one in the 2011 flare) that could be detected by the CTA northern array (for different values of B and $\Gamma_{e}$). For each of those, we performed 5000 observation simulations of the nebula in both steady and flaring state, and applied chi-square tests between the excess distributions as explained above. Instead of the 10h considered above, we used different observation times, ranging from 0.01\,h to 500\,h (in adaptative bins varying from 0.01\,h to 1\,h in size), and calculate the minimum time to detect a significant variation with respect to the steady state. To consider the varying observation time properly, we used the IRFs optimized for an exposure time of 50\,h, 5\,h, and 0.5\,h and re-scaled the total energy in electrons above 1\,TeV ($W_{\rm{e}}$) from the one obtained in 10\,h.

Finally, we have not considered in our simulations the effect of the systematic errors, which should be in principle the main limitation to measure variability over the nebula baseline flux. The many uncertainties on the CTA final systematics prevent us from estimating its quantitative effect on the simulations. Therefore, the results of this work should be considered as an optimistic case.

\section{Results}
\label{sec3}

\subsection{Application to the Crab 2011 gamma-ray flare}

\begin{figure*}
  \centering
   \includegraphics[width=0.45\textwidth]{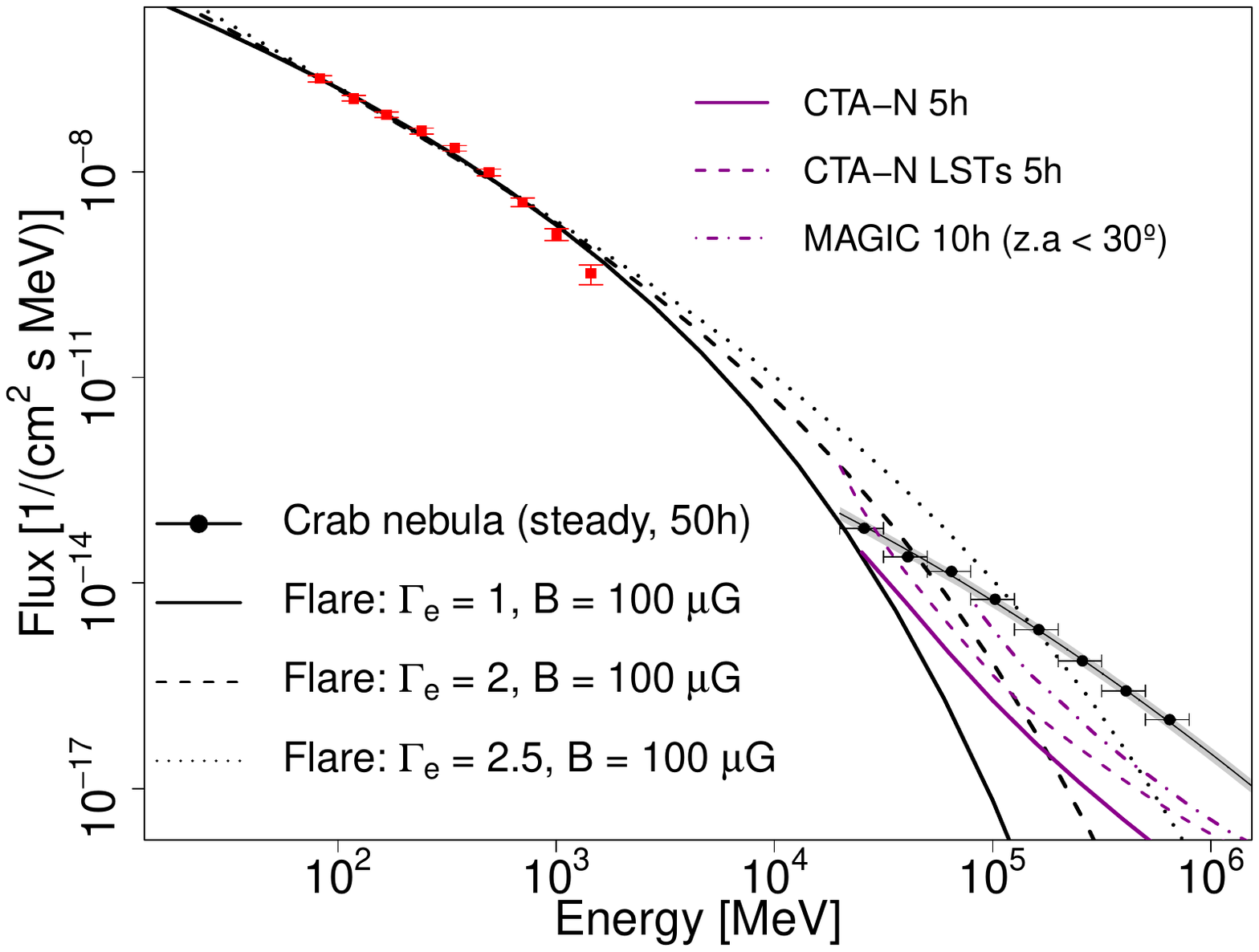}
   \includegraphics[width=0.45\textwidth]{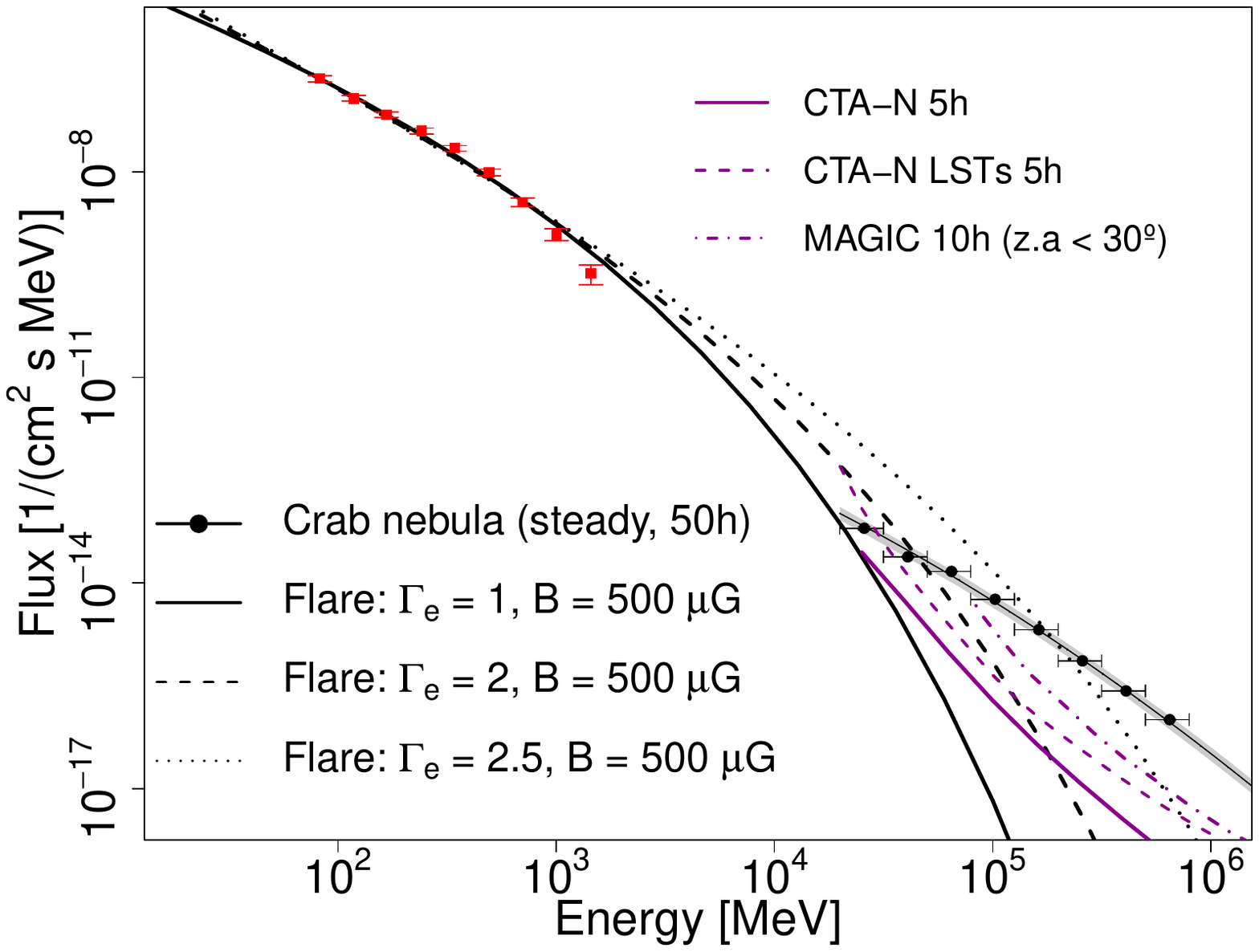}
  \caption{The synchrotron emission for flares  (fitted to the 2011 April flare spectrum) of different indices ($\Gamma_{e} = 1, 2, 2.5$) and magnetic fields ($B = 100\ \mu$G in the left panel, and $500\ \mu$G in the right one) together with the simulations of the Crab nebula steady spectrum (in black line with dots), with the $3\sigma$ region noted (black shaded area, with only statistical errors). The data recorded by \emph{Fermi} of the 2011 flare is plotted in red squares. The magenta solid and dashed lines correspond to the sensitivities of the CTA northern array and if considering only its four Large Size Telescopes (for 5\,h of observation time), respectively (see the instrument response functions\protect\footnote[1]{}). The magenta dot-dashed line corresponds to the sensitivity of MAGIC stereo system at low zenith angle ($< 30\degr{}$) \citep[see Tables A.5 and A.6 in][]{2016Aleksic}.}
  \label{fig:Synchrotronflaremodels}
\end{figure*}

\begin{figure*}
\centering
  \includegraphics[width=0.45\textwidth]{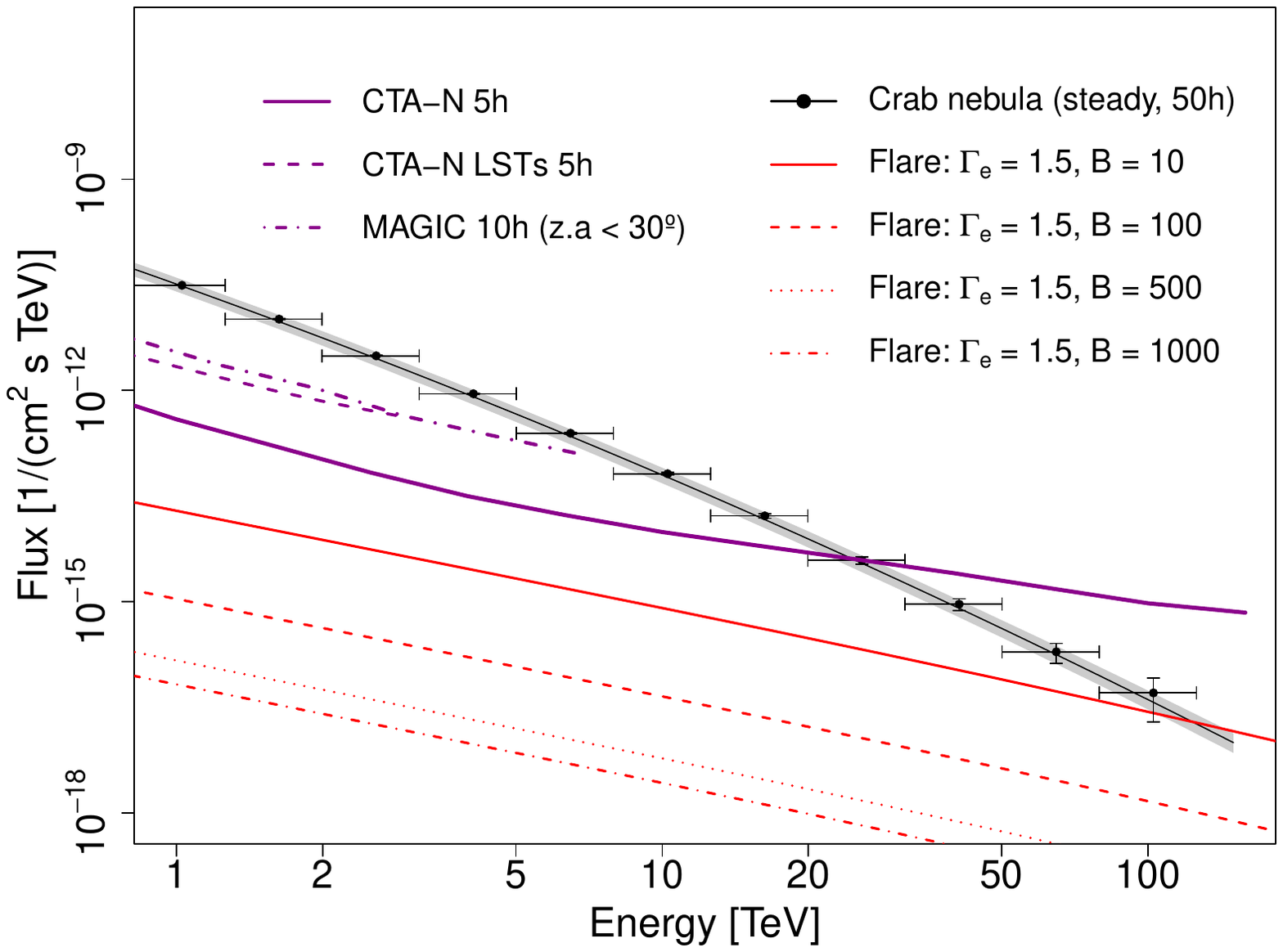}
  \includegraphics[width=0.45\textwidth]{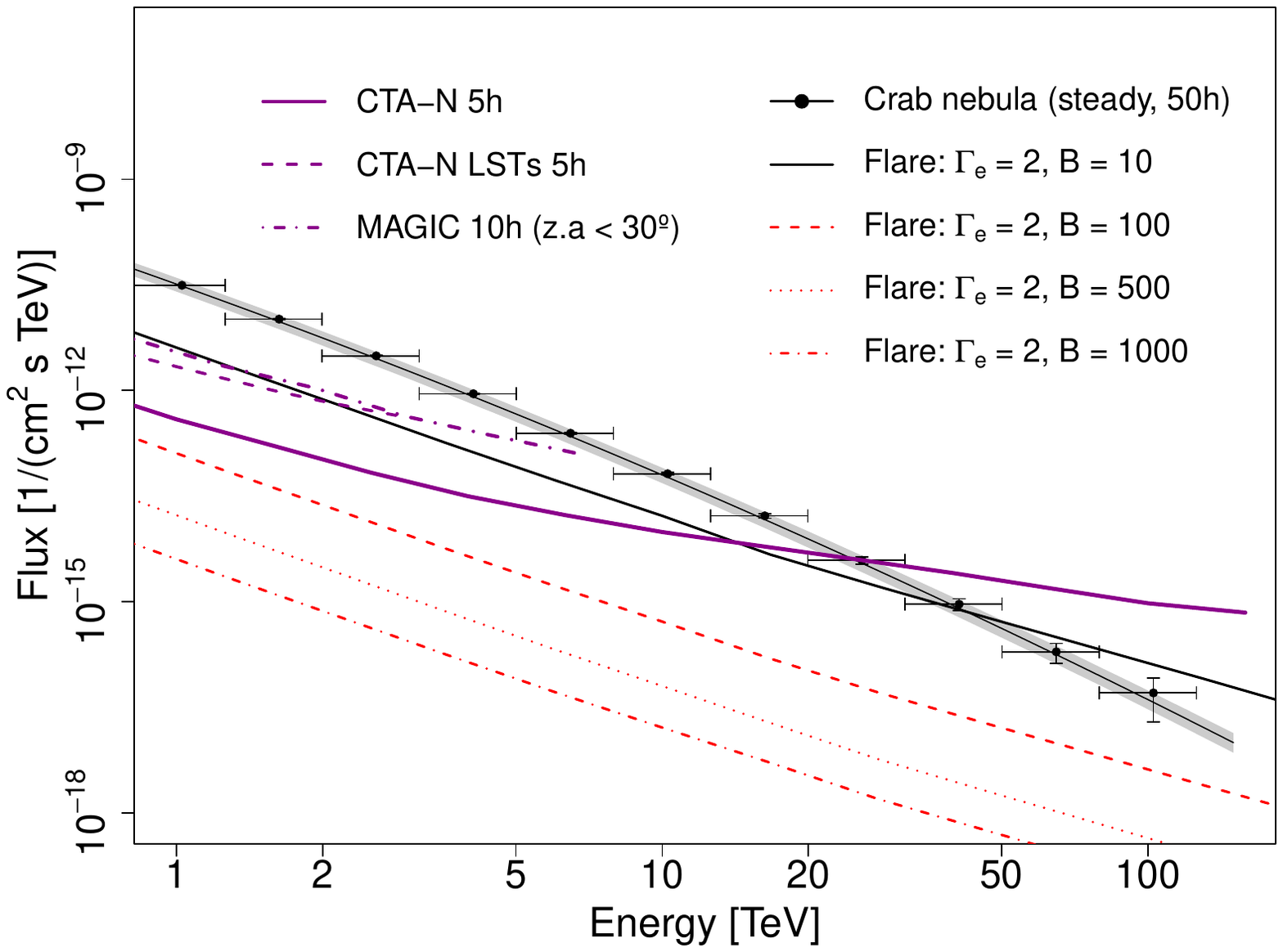}
  \caption{IC emission for different models of flares (fitted to the 2011 April flare spectrum, in black line if the flare is detected in the TeV regime, and in red line otherwise) with $ \Gamma_{\rm{e}} =1.5$ (left) and $ \Gamma_{\rm{e}} =2$ (right) on top of the simulations of the Crab nebula steady spectrum (in black line with dots), with the $3\sigma$ region noted (black shaded area, with only statistical errors). The SSC emission in a region of $2.8 \times 10^{-4}$ pc in size is taken into account. The sensitivity curves (in magenta) are defined in Figure \ref{fig:Synchrotronflaremodels}.}
  \label{fig:ICflaremodels}
\end{figure*}

The general effect of the magnetic field and $\Gamma_{\rm{e}}$ on the expected $\gamma$-ray spectra of the flares is summarised in Fig. \ref{fig:Synchrotronflaremodels} (for the synchrotron part), and in Fig. \ref{fig:ICflaremodels} (for the IC one). The first figure focuses on the sub-100\,GeV region. The synchrotron spectrum in the GeV regime does not depend significantly on the magnetic field, due to the fitting of the synchrotron emission to the LAT data, which constrains strongly the synchrotron flux level above hundreds of MeV. We show (see Table \ref{tab:Flaredetectionlo.} and Fig. \ref{fig:Synchrotronflaremodels}) that CTA should be able to detect a flare with similar flux to the one observed in 2011 in all cases tested. In fact, our simulations prove that the detection should be possible in less than one hour for $\Gamma_{\rm{e}} > 1.0$ below 200\,GeV, representing a variation of integral flux in the GeV regime stronger than 20 per cent of the nebula steady flux. The detection was achieved for stepper spectra ($\Gamma_{\rm{e}} > 2.0$) in less than an hour, even at higher energies (up to 300\,GeV). This is possible due to the large flux expected on the GeV regime when the threshold is low enough. For instance, the predicted excess (in counts) for the steady nebula (with 10\,h of CTA northern array) from 20\,GeV to 120\,GeV, is $\sim 2$ times the excess predicted from 1.25\,TeV to 50\,TeV. To emphasize the effect of detectability of the synchrotron part, given the high flux expected as the energy threshold decreases, we also showed in Fig. \ref{fig:Synchrotronflaremodels} the sensitivity of CTA using only 4 Large Size Telescopes (LSTs) in the North site, and the one of the MAGIC telescopes \citep{2016Aleksic}. For bright flares, the large CTA-LST sub-array should be sufficient to impose strong constrain on the synchrotron tail. Even for current instruments like MAGIC (see Fig. \ref{fig:Synchrotronflaremodels}, magenta dot-dashed line), prospects are optimistic, allowing a good determination of the particle spectrum if the right observation conditions are met, i.e., for a flux level similar to the one observed in the April 2011 flare, observations at low zenith angle (E < 300 GeV) may have resulted in a detectable enhancement of flux for a soft particle index ($\Gamma > 2$). 


On the contrary, the magnetic field has a strong effect on the TeV IC component (see Table \ref{tab:Flaredetectionhi.} and Fig. \ref{fig:ICflaremodels}), and strongly suppresses the flare contribution for very large intensities. Opposite to the sub-100 GeV regime, soft electron spectra favour the production of GeV-TeV photons, by promoting lower energy electrons which provide a larger SSC photon target field. 
From the simulations, we derived that only certain combinations of parameters characterising the flare emission result on a detectable variability over the total flux. If the power-law index of emitting particles is hard, $\Gamma_{\rm e} < 2.2$, then the IC emission from this population is detectable only for very weak magnetic fields, $\lesssim 50$ $\upmu$G. For softer electron distributions, i.e. $\Gamma_{\rm e} > 2.5$, realizations with magnetic fields as large as 1\,mG result in detectable levels of flux. 
  
\begin{figure}
  \centering
  \includegraphics[width=0.5\textwidth]{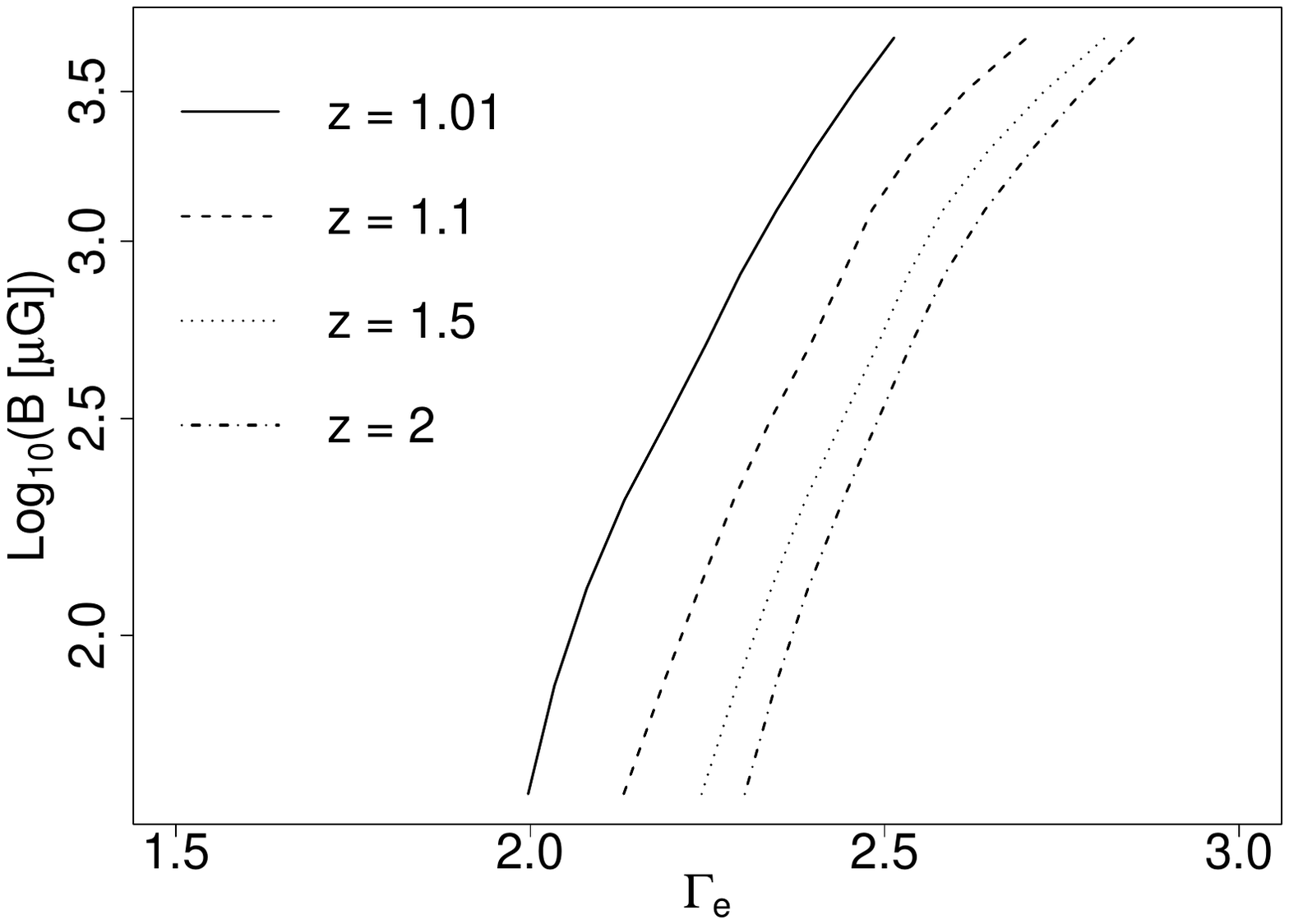}\\
  \caption{The enhancement of integral flux above 1\,TeV (in Eq. \ref{eq:zparameter}) for the different models of flare (fitted to the 2011 April flare spectrum), taking into account the SSC.}
  \label{fig:MAGIClikeflareintegralfluxes}
\end{figure}

The observation times listed on Tables \ref{tab:Flaredetectionlo.} and \ref{tab:Flaredetectionhi.} are obtained by evaluating the excess observed over the contribution of the nebula (see Section \ref{sec2.4}). We attempted to reconstruct the spectral shape of the flare contribution, but only in a handful of cases a clear deviation with respect to the nebula steady spectrum was measured. For example, for $\Gamma_{\rm{e}} = 2.5$ and $\rm{B} = 100$ $\upmu$G, a significant difference of spectral indices of $0.66 \pm 0.03$ was measured below 300\,GeV, with respect to the index measured in steady state for the GeV regime ($2.33\pm 0.03$), and with the break located at an energy of $\sim 300$\,GeV. This effect is also present in the TeV regime (at a $3\sigma$ level), for which a significant variation in the spectral index ($0.46 \pm 0.03$, with respect to $2.75\pm0.03$) is observed. Figure \ref{fig:MAGIClikeflareintegralfluxes} shows the enhancement factor as defined in Eq. \ref{eq:zparameter}, applied to the flux of the flaring state and steady one above 1\,TeV. 

It is also interesting to note that none of the performed simulations results on a flux level above the nebula at energies below 1\,MeV. Thus, low spatial resolution (> arcmin scale) instruments at lower energies cannot resolve the morphology of the nebula (see Fig. \ref{fig:Flareexample}), preventing the detection of counterparts. However, this may not be the case if a power-law spectral shape cannot characterize the particle spectrum.

%
\begin{table}
\centering
\footnotesize
\caption{Observation time (in hours) necessary to detect different models of flares (fitted to the 2011 April flare spectrum) at energies from 20\,GeV to 120\,GeV with the chi-square test applied to the excess distributions (at 99\% CL). Note that we assume that Crab is in flaring state during the entire time of observation.}
\label{tab:Flaredetectionlo.}
\begin{tabular}{clclclclc}
\hline
$\Gamma_{e}$ & $1.0$ & $1.5$ & $2.0$ & $2.5$ \\
\hline
Time [h] & $0.35$ & $0.14$  & $0.02$ & $<< 1$ \\
\hline
\hline
\end{tabular}
\end{table}

\begin{table}
\centering
\footnotesize
\caption{Observation time (in hours) necessary to detect different models of flares (fitted to the 2011 April flare spectrum) at energies from 1.25\,TeV to 50\,TeV, obtained as Table \ref{tab:Flaredetectionlo.}. We assume that Crab is in flaring state during the entire time of observation. The models indicated with asterisks imply $W_{\rm{e}} > 5\times10^{43}$ erg, if the duration of the flare is the one indicated. The Crab steady nebula was detected in 3.2\,h at $5\sigma$ ($\sim 9\sigma$ in 10\,h).}
\label{tab:Flaredetectionhi.}
\begin{tabular}{clclclc}
\hline
\diagbox{B[$\upmu$G]}{$\Gamma_{\rm{e}}$} & $1.5$ & $1.8$ & $2.1$ & $2.5$  \\
\hline
$1000$ & $> 500$ & $> 500$ & $> 500$ & $8.0$ \\
$500$ & $> 500$ & $> 500$ & $> 500$ & $0.8$\\
$100$ & $> 500$ & $> 500$ & $160^{*}$ & $0.02$\\
$50$ & $> 500$ & $> 500$ & $40^{*}$ & $<< 1$\\
$10$ & $> 500$ & $93^{*}$ & $0.3^{*}$ & $<< 1$\\
\hline
\hline
\end{tabular}
\end{table}

However, the numbers above do not consider the limitations regarding the total energy available in the system (limited to $\sim 5 \times 10^{43}$ erg, see Section \ref{sec2.1}). If no re-acceleration is taken into account (or any other way to provide a larger energy budget), only some combinations of magnetic field and spectral indices are therefore possible. For $\Gamma_{\rm e} > 2.2$, the models with magnetic field below $\rm{B} \sim 500$ $\upmu$G are ruled out ($W_{\rm{e}} > 5 \times 10^{43}$ erg). This value has to be increased to 1\,mG when considering softer indices, $\Gamma_{\rm e} > 2.5$. For spectral indices of the order of $\Gamma_{\rm e} > 1$, only magnetic fields stronger than $\sim 150$ $\upmu$G keep the total energy below the limit. This can be seen in Fig. \ref{fig:flare2011}, in which the shaded area mark the, in principle, forbidden values of B and $\Gamma_{\rm e}$.

\subsection{Application to the Crab 2013 gamma-ray flare}

We applied the simulation scheme in Sections \ref{sec2.3} and \ref{sec2.4} to fit the \emph{Fermi} data of the 2013 flare from 80\,MeV to 1\,GeV. We considered the same parameters space as before, with the magnetic field strength in the range of 10\,$\mu$G to 5\,mG and $\Gamma_{\rm{e}}$ ranging from 1 to 3. For each model of flare simulated we computed the integral flux above 1\,TeV, fitting the total IC (with the CMB, NIR and FIR photon fields) and taking into account the SSC in a region of $1.7 \times 10^{-4}$ pc in size, and compared it to the integral flux of the simulations of the steady nebula. Then, the flux enhancement (see Eq. \ref{eq:zparameter}) for the different models of flares was compared to an upper limit similar to ones reported by H.E.S.S. and VERITAS, constraining the accelerating magnetic field and the particle index of the electron population.    

\begin{figure}
\centering
  \includegraphics[width=0.45\textwidth]{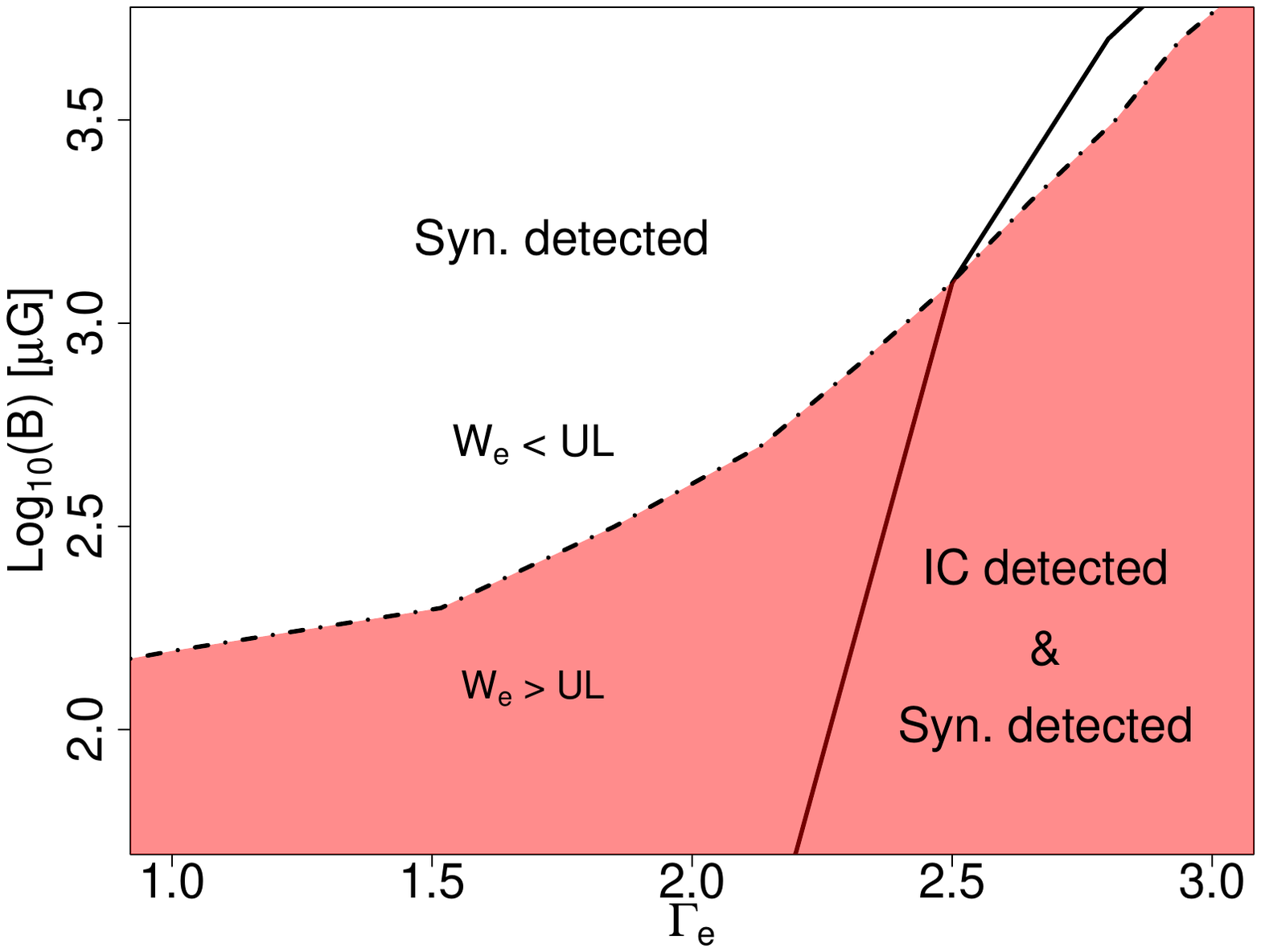}
\caption{Results of the observation simulations of the flaring nebula, for the different models of flares, fitted to the \emph{Fermi} data of the April 2011 flare. The models located below the solid line would be detected both in the TeV and GeV regimes (i.e., from 1.25\,TeV to 50\,TeV of energy, and below 200\,MeV), if CTA would have observed, while the models above the solid line are detected only in the GeV regime. However, the models below the dash-dotted line (red shaded area) require an energy in electrons $W_{\rm{e}} > 5\times 10^{43}$ erg.}
\label{fig:flare2011}
\end{figure}

\begin{figure}
\centering
\includegraphics[width=0.45\textwidth]{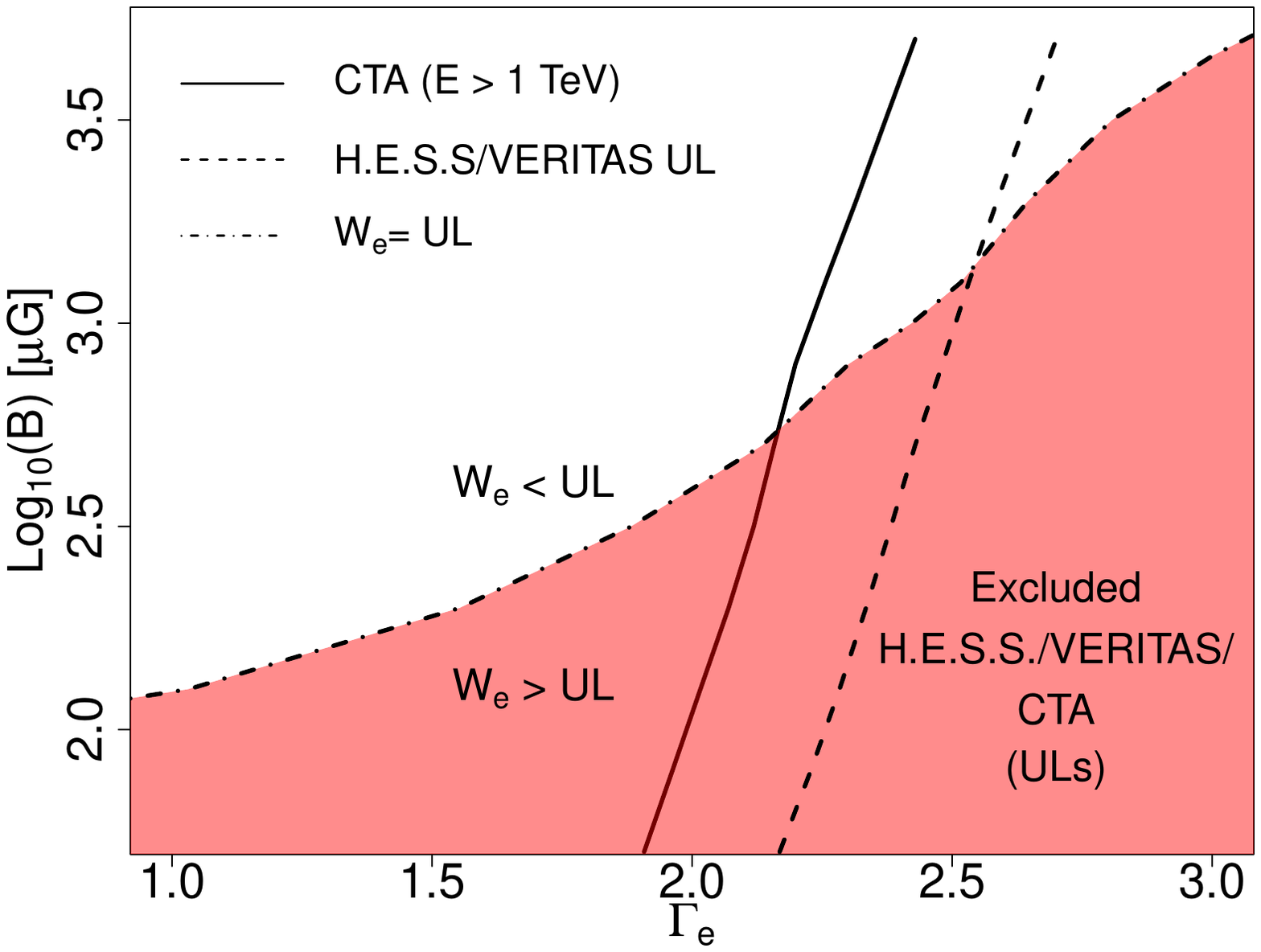}
\caption{Results of the observation simulations of the flaring nebula, for the different models of flares fitted to the \emph{Fermi} data of the March 2013 flare, compared to the upper limits established by H.E.S.S. and VERITAS observations. The models below the dashed line correspond to an enhancement of integral flux above 1\,TeV larger than a 60\% (i.e Z > 1.6 in equation \ref{eq:zparameter}). Therefore, these models violate the upper limits established by the H.E.S.S. and VERITAS observations of the 2013 flare. The exclusion region (for the integral flux upper limit above 1\,TeV) extends up to the solid line (Z > 1.01, with only statistical errors) supposing the flare is not detected with 50\,h of CTA northern array. The models below the dashed-dotted line (red shaded area) violate the upper limit estimated for the total energy in electrons.}
\label{fig:flare2013}
\end{figure}


The flares are detected in the GeV regime for $\Gamma_{e} > 1.4$, in 10\,h of observation time, implying an enhancement of flux (and counts, in the GeV regime) above a 4 per cent with respect to the steady nebula. 
The results in the TeV regime are summarised in Fig. \ref{fig:flare2013}. The shaded area shows the exclusion region as on Fig. \ref{fig:flare2011}, due to the nebula energy budget. The CTA exclusion region (solid line) improves with respect to the one constrained by the H.E.S.S. and VERITAS observations (dashed line). It is interesting to note, that if the energy available is the one provided by Eq. \ref{eq:recon_Lorentz}, only a few combination of B and $\Gamma_{\rm e}$ results on a detectable flux, being this restriction more constraining than the one imposed by observations with CTA or the upper limits from the H.E.S.S. and VERITAS telescopes.

\subsection{A power-law synchrotron spectrum in the GeV regime}
\label{sec:3.3}

In some theoretical approaches, the resulting emission beyond the maximum of the radiation might follow a power-law extending to high energies, instead of the standard exponential cutoff \citep[as expected in turbulent magnetic field, see, e.g.,][]{Fleishman2006,Kelner2013}. To evaluate this possibility, we fit the \emph{Fermi} data above 400\,MeV of the April 2011 flare with a power-law photon spectrum and investigate the detectability in the GeV/TeV regime using the same approach described in Section \ref{sec2.4}. The fit to the LAT data results on a spectral photon index of $3.14 \pm 0.13$ (at $1\sigma$) and a flux at 850\,MeV of $(1.33 \pm 0.07)\times 10^{-9}$ $\rm{cm}^{-2}\ \rm{s}^{-1}\ \rm{MeV}^{-1}$ (with a reduced $\chi^{2} = 1.4$ for two degrees of freedom). Note that the photon spectrum of the flare in the GeV regime is not derived, in this case, from fitting the spectrum of an electron particle population.

The simulations of the flare fitted to the data with $\Gamma_{\rm{Jit}} = 3$ showed that the flare is always detected below 500\,GeV (in 10\,h), with the chi-square tests applied to the excess distribution. This model corresponded to an increase of integral flux of about $\sim 70$\% in the GeV regime. 
We also tested a softer model with $\Gamma_{\rm{Jit}} = 3.5$, which produces also a detectable emission (in the same time) below 200\,GeV of energy, with an increase of integral flux (and excess counts) of $\sim 9$\% in the GeV regime. 
For harder spectrum the scenario is optimal. Using a $\Gamma_{\rm{Jit}} = 2.5$, the flare emission above the nebula can be observed up to TeV energies in 10\,h. The integral flux is up to $\sim 6$ ($\sim 2$) times larger than in steady state, in the GeV (TeV) regime. The enhancement of integral flux above 1\,TeV, in this case, is therefore above the upper limit established by H.E.S.S and VERITAS (see Section \ref{sec2.2}).

Note that the simulations of the IC flares depend on the synchrotron emission of a fitted particle population (characterized by an exponential cutoff power-law spectrum, see Section \ref{sec2.3}), which would not reproduce in any case a power-law spectrum emission above 400\,MeV as described above. Therefore, the simulations do not allow us to compute the IC emission or the particle spectrum in this scenario, and as a result, the energy in electrons ($\rm{W}_{e}$) either.

\section{Discussion}
\label{sec4}

The rapid MeV-GeV flares cannot be easily explained using the standard approach of PWN theory. In such, below few tens of GeV, the emission is believed to be dominated by the high-energy tail of the synchrotron emission produced by leptons in the shocked pulsar wind. This emission is limited, in an ideal MHD outflow, by the synchrotron burn-off limit, which in the plasma co-moving frame the peak of spectral energy distribution is below  $\hbar \omega\ma < 200$~MeV \citep[assuming that the acceleration and radiation sites are co-located, see, e.g.,][]{1983MNRAS.205..593G}. The steady spectrum measured in the Crab nebula seems to agree with this limit. To explain the variable emission that peaks above this limiting energy, a second component has to be invoked, that emerges at a few \(100\)~MeV with a flux exceeding the nebula one. The origin of this second component however is unclear, and furthermore, the gamma-ray emission seems to be inconsistent with traditional synchrotron or IC mechanisms: while the fast variability timescale robustly rules the IC scenario out, the large peaking energy excludes the synchrotron scenario in the ideal MHD setup.

Since the discovery of the first flare \citep{Tavani2011}, several interpretations have been put forward  to explain the Crab flares. Within the limits of an ideal MHD outflow, a relativistic-moving acceleration site \citep[see, e.g.,][]{2011MNRAS.414.2229B,2012MNRAS.427.1497L,2018JPlPh..84b6301L} and/or a radiation mechanism other than synchrotron \citep[i.e., Jitter radiation as in][]{2013ApJ...763..131T} could in principle account for the observed emission.
Alternatively, if the MHD assumptions are violated in some region, for example, due to reconnection of the magnetic field, an electric field component parallel to the magnetic field can be generated, resulting on a beam of high energy particles that will move along the unscreened electric field. These accelerated particles eventually escape from the region of acceleration and can be deflected by the average magnetic field. If the particles have super-critical energy, they should emit a substantial part of their energy over a time short compared to the girorotation \citep[see, e.g.,][]{2012ApJ...754L..33C}. A somewhat similar situation can be realized if a beam of particles is accelerated in a low pair loading wind \citep{2017PhRvL.119u1101K}.

In the following, we discuss the constraints derived from our simulations to the different approaches, when possible, and offer different possiblities and prospects for CTA.

\subsection{Relativistically Moving MHD Outflow}

\label{sec:relativ_outflows}

In the first class of scenarios (i.e. those that invoke relativistic MHD outflows), the emission site can be described as a region of size $R'$, moving with a significant bulk Lorentz factor \(\Gamma\), such that the Doppler boosting factor is $\delta_{10}\sim1$ (here $\delta_{10} = 0.1/(\Gamma(1 - \beta \cos \theta))$). In such a case, the strength of the comoving magnetic field $B'$ should be high enough to satisfy the Hillas criteron:
\begin{equation}
      R' > 3\times 10^{15}{E'}_{\rm PeV}{B'}^{-1}_{{\rm mG}}\, {\rm cm} 
\label{eq:hillascriterion_1}
\end{equation}
where $E'_{\rm PeV}$ is the maximum comoving electron energy (in PeVs). Thus, the following condition must be satisfied:
\begin{equation}
     E_{peak} = 600 {E'}_{\rm PeV}^{2}{B'}_{{\rm mG}}\ \delta_{10}\, {\rm MeV} 
\end{equation}
 Also, if the synchrotron cooling time is responsible for the variability:
 \begin{equation}
     t_{syn} = 4\times 10^{4}{E'}_{\rm PeV}^{-1}{B'}_{{\rm mG}}^{-2}\ \delta_{10}^{-1}\, {\rm s} 
\end{equation}
Combining the expressions above we obtain:

\begin{equation}
E'_{\rm PeV} > \left( \frac{E_{peak}}{600\,{\rm MeV}} \right)^{2/3} \left( \frac{t_{\rm var}}{4\times 10^{4}\, {\rm s}} \right)^{1/3}\delta_{10}^{-1/3}
\label{eq:partenergyinblob}
\end{equation}
and
\begin{equation}
B'_{\rm mG} < \left( \frac{E_{peak}}{600\,{\rm MeV}} \right)^{-1/3} \left( \frac{t_{\rm var}}{4\times 10^{4}\, {\rm s}} \right)^{-2/3}\delta_{10}^{-1/3}
\label{eq:magneticfield}
\end{equation}
In the expression above $E_{peak}$ refers to the peak photon energy in the observer frame and $t_{\rm var}$ to the flare variability time, with $t_{\rm var} > t_{\rm syn}$.

Using Eqs.~\eqref{eq:partenergyinblob} and \eqref{eq:magneticfield}
in Eq.~\eqref{eq:hillascriterion_1}, one can retrieve the blob size in terms of the peak energy  $E_{peak}$ and variability time $t_{var}$:
\begin{equation}
    R' > 3\times 10^{15} \left( \frac{E_{peak}}{600\,{\rm MeV}} \right) \left( \frac{t_{\rm var}}{4\times 10^{4}\, {\rm s}} \right)\, {\rm cm}
\label{eq:hillascriterion_2}
\end{equation}

The size of the nebula limits physically the maximum of the blob size to $\sim 3.4\times 10^{18}$\,cm. 

Using the approximations above, we can estimate the minimum and maximum values of the Doppler boosting factor ($\delta_{10}$), by comparing with our simulations. For the different values of magnetic field sampled, we obtain $\delta_{10_\rm max}$ using Eq. \ref{eq:magneticfield}. We derive then from our simulations $\rm{E'_{PeV}}$ in the comoving frame of reference, and obtain the size of the blob using Eq. \ref{eq:hillascriterion_1}. To check if this approximation is consistent with our data, the parameters resulting from the simulations ($E'_{\rm PeV}$ and \(R'\)) are compared with the ones obtained with Eq. \ref{eq:partenergyinblob} and \ref{eq:hillascriterion_2}, respectively.

\begin{figure}
  \centering
\includegraphics[width=0.5\textwidth]{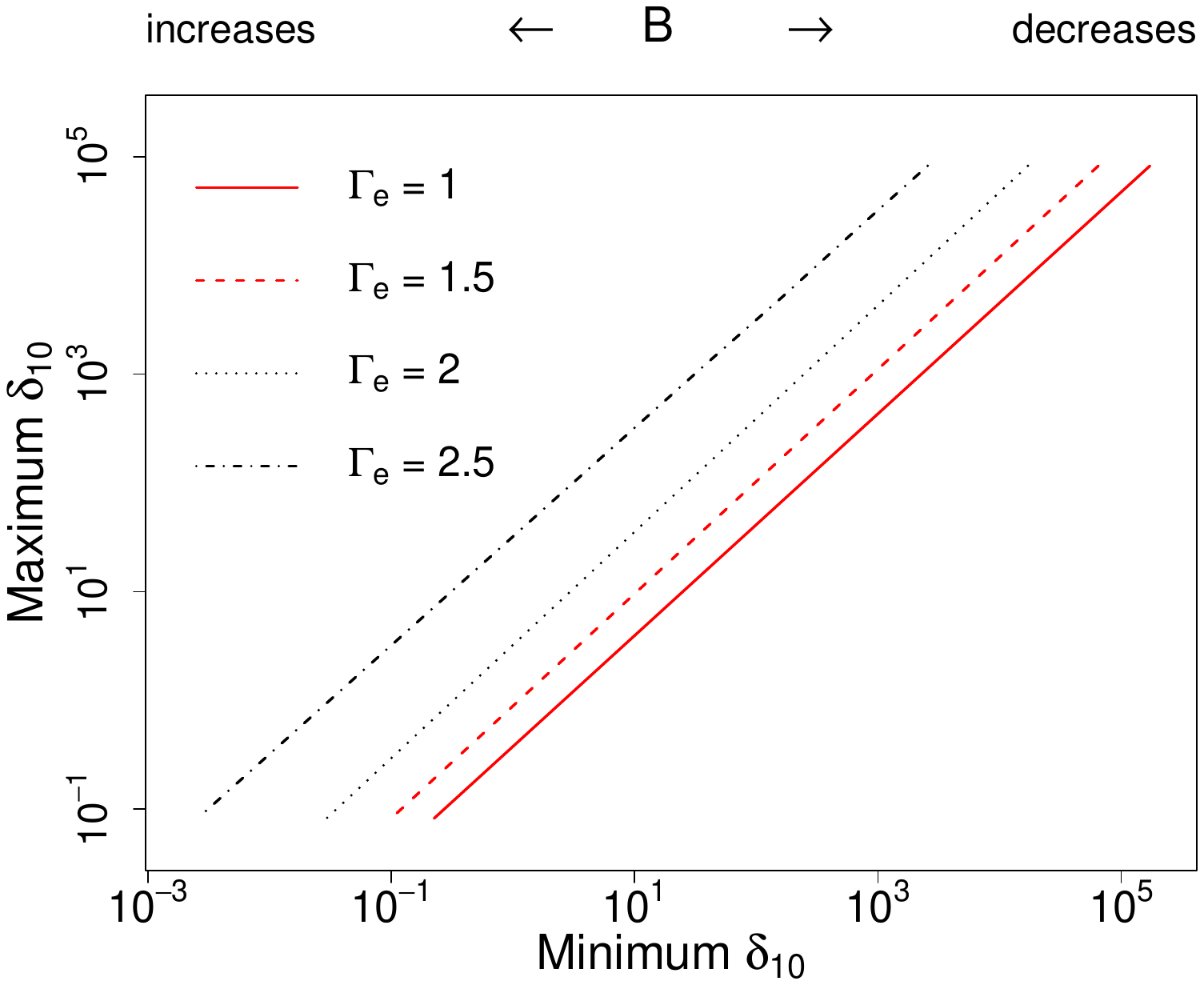}\\
  \caption{Doppler factor lower and upper limits from Eq. \ref{eq:partenergyinblob} and \ref{eq:magneticfield}. The simulations plotted in red are not consistent with the relativistic moving outflow scenario for any Doppler factor, since the cutoff energy of the particle spectrum ($\rm{E'_{PeV}}$) fitted  in these cases is too small for the Doppler boosting factor upper limit derived with Eq. \ref{eq:magneticfield}.}  
  \label{fig:minmaxsigma10}
\end{figure}

We conclude that $E'_{\rm PeV}$ is consistent with the relativistic moving outflow scenario if $\Gamma_{e} > 1.5$ (within a 10\% of error), allowing a certain range of $\delta_{10}$. The upper and lower limit of $\delta_{10}$ must satisfy in each case the inequalities in Eq. \ref{eq:partenergyinblob} and \ref{eq:magneticfield}. Figure \ref{fig:minmaxsigma10} shows the limits on the Doppler boosting factor derived from the comparison with the simulations. A detection on the TeV regime would then constrain the Doppler factor to a very limited range. For instance, models detectable in the TeV regime, with mG magnetic fields (and $\rm{W}_{e} < 5 \times 10^{43}$ erg, see Figure \ref{fig:flare2011}), are only feasible for Doppler factors ranging from $0.026$ to $51.9$.

Similarly, the minimum size of the blob computed from simulations is consistent with Eq. \ref{eq:hillascriterion_2} ($R' > 4.5 \times 10^{14}$ cm), for $\Gamma_{e} > 1.5$, resulting at most $1.8 \times 10^{15}$ cm for $\Gamma_{e} = 3$, still $\sim 2000$ times smaller than the size of the nebula. 

\subsection{Non-synchrotron scenarios}

Alternatively to the relativistic motion case, the flare emission can be also generated in an ideal MHD outflow if the radiation mechanism is not synchrotron. Jitter radiation, i.e., a modification of the classical synchrotron radiation for the case turbulent magnetic field \citep[see][and references therein]{Kelner2013}, was suggested as a possible channel for the flares \citep{2012Bykov,2013ApJ...763..131T}.
The jitter mechanism is characterized by the same cooling time as the synchrotron one (for the same mean magnetic field strength) but the peak energy is increased by a factor $R_{g}/(\gamma \lambda_{turb})$, where $R_{g}$, $\gamma$, and $\lambda_{turb}$ are giro-radius for the mean magnetic field strength, particle Lorentz factor, and turbulent field correlation length.

If the correlation length is small enough (i.e smaller than photon formation length), one would expect a shift on the total spectral energy distribution towards higher energies, facilitating the detection of the flare at the lower energies of CTA. In this case, the resulting spectrum can differ strongly from synchrotron: not only the maximum of the spectrum will be shifted to larger energies, but also the shape of the spectrum above the peak would follow a power-law distribution \citep{Fleishman2006,Kelner2013}, being the emerging index related to that of the spectrum of the particle distribution and the magnetic turbulence. A power-law high-energy tail is also expected if the correlation length of the magnetic field is comparable to the giro-radius \citep[however in this case the peak maximum would remain unchanged, see][]{Kelner2013}. 

We tested this scenario by fitting the flare spectrum above 400\,MeV observed in the 2011 flare, with a power-law function (see Section \ref{sec:3.3}). The spectrum is well-described with a photon index of $\Gamma_{\rm{Jit}}\sim$3. Our simulations show that such spectral shape would result on a clear detection with CTA, with a substantial increase of the integral flux in the sub-100 GeV regime (up to $\sim$70\%), if the power-law function continues with such index to high energies. Softer indices up to 3.5 would also result on a sizeable excess above the nebula emission below 200\,GeV in 10\,h, while a harder index ($\Gamma_{\rm{Jit}} \sim 2.5$) could be observed at TeVs with an enhancement of flux similar to or larger than the upper limit set by H.E.S.S. and VERITAS to the 2013 flare.

If the jitter mechanism is therefore behind the origin of the Crab flares, the large excess expected below a few hundreds of GeV would result in a clear detection with CTA, of bright, but also moderated, flares, possible to achieve even in early stages of the construction phase.

\subsection{Non-MHD scenarios}

\begin{figure}
  \centering
  \includegraphics[width=0.5\textwidth]{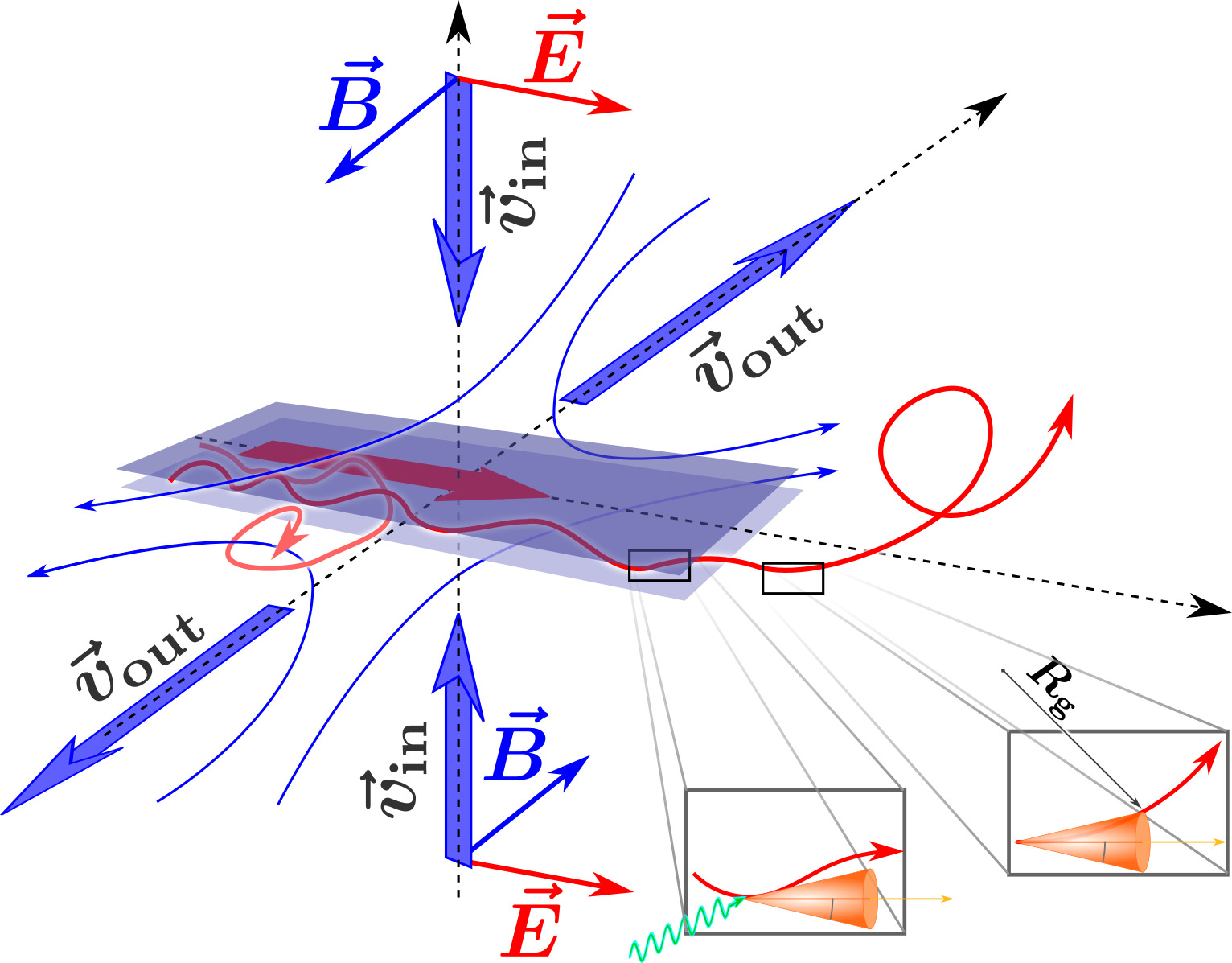}\\
  \caption{Sketch of Parker-type reconnection region. Two magnetized flows with oppositely directed magnetic field \(B\) converge with velocity \(V_{\rm in}\) into the reconnection region schematically shown with blue planes. The electric field \(E\), which is co-directed in the converging flow, appears unscreened (shown with thick red arrow) in the reconnection region. Particles propagating in the reconnection region (shown with red lines) can be accelerated to high energies before their escape (plasma escaping from the reconnection region is shown with \(v_{\rm out}\). Accelerated particles can produce IC emission or synchrotron emission in the magnetic field outside the reconnection region (radiation processes are illustrated in in-figure boxes).}  \label{fig:relativistically_moving_blobs}
\end{figure}

The third possibility discussed in the literature refers to an alternative mechanism, in which in certain regions of the nebula, magnetic field lines of different orientation draw near each other and create a reconnection layer (see Fig. \ref{fig:relativistically_moving_blobs}). 

There is  a number of processes that are activated by magnetic reconnection. First of all, the flow configuration is such that the initially highly magnetized flow that enter  the reconnection layer (this motion is vertical in Fig. \ref{fig:relativistically_moving_blobs}) gets ejected from the reconnection layer. The outlet MHD flow has a high internal energy and might propagate with large bulk Lorentz factors (see for example in the case of Petchek-type relativistic reconnection, \citealt{2009ApJ...696.1385Z}). Non-thermal processes operating in such outflow operate in the same way as discussed in Sec.~\ref{sec:relativ_outflows}. If the guiding magnetic field is small, then the magnetization of the flow should be small favoring higher IC fluxes. This, however, requires an almost perfect alignment of reconnecting magnetic field, which seems to be a significantly less probable realization than one with not a strictly anti-parallel magnetic field. In this case the guiding magnetic field is comparable to the reconnecting field and  the flow remains highly magnetized. In this case, the IC emission is significantly suppressed. Typically, one adopts mG magnetic field in reconnection scenarios \citep[see, e.g.,][]{2012ApJ...754L..33C}, which makes very hard detection of the counterpart IC emission. 

If particles are directly accelerated by magnetic field in the reconnection layer (along the thick red arrow in Fig.~\ref{fig:relativistically_moving_blobs}), then during the acceleration process they effectively reside in a region with weaker magnetic field, which suppresses the synchrotron cooling, but the particles still produce some IC emission. The acceleration time in the reconnection layer, \(t_{\rm rl,acc}\simeq E_e/ev_{\rm in} B\), is longer by a factor of approximately \(c/v_{\rm in}\) than the synchrotron cooling time in magnetic field of \(B\) (which, given the location of the synchrotron peak at a few hundreds of MeV, should be about \(t_{\rm syn}\simeq E/eBc\)). Thus, if the reconnection operates in a non-relativistic regime, \(v_{\rm in}\ll c\), then the emission from the reconnection layer may provide the dominant IC contribution. As in this case the particles produce IC and synchrotron emission at very different parts of their trajectories, these two components may have quite different beaming patterns. It is interesting to note, that in principle one could expect {\it orphan} flares in the TeV regime. That opens exciting prospects when observing bright PWNe in which the spectrum can be measured with good statistics. 

The many uncertainties on these models, however, prevent us to make quantitative estimation of the expected radiation based on our simulations.

\section{Conclusions}
\label{sec5}

The simulations performed open exciting prospects for CTA, in particular for the LSTs, that sample with best sensitivity the GeV energy range. We showed that the detection of flares in the Crab nebula at tens of GeV with CTA, can be achieved in the case of synchrotron radiation with hard particle indices ($\Gamma_{e} = 1.5 - 2.0$) and magnetic fields similar to or stronger than the estimated average magnetic field in the nebula ($B \gtrsim 150 \mu$G, implying an energy in TeV electrons for the parent particle population smaller than $5 \times 10^{43}$ erg, see Figure \ref{fig:Synchrotronflaremodels}). It should be noted, that the formulation of the synchrotron radiation used in NAIMA is somehow conservative. Several works \citep{2019ApJ...887..181D,2020arXiv200300927K} have demonstrated that the spectrum of synchrotron radiation can significantly deviate from the one obtained in random magnetic fields \citep{2010PhRvD..82d3002A}, and extend into the very high energy regime. Observation with Cherenkov telescopes of Crab flares, ensuring an optimal energy threshold, provide a unique opportunity to investigate the magnetic structure in astrophysical environments with great detail. Furthermore, we compared the expected emission from the synchrotron tail with the sensitivity curves for 4 CTA-LSTs and the MAGIC experiment (which can observe the Crab Nebula with low energy threshold). We show that observations of bright flares, similar to the one in 2011, could already provide strong constrains, even in the most pessimistic synchrotron approach considered here. 

The IC component faces, in general, more complications: in one hand the variability has to be observed over the overwhelming nebular background, on the other, low magnetic fields should be involved to boost the IC emission. The latest, even if theoretically possible, seems unlikely when considering the fast variations observed in the Crab Nebula light curve during the flares. We found additionally an important drawback: to achieve a detectable TeV flux level in most of the cases explored, the total energy required should exceed the one contained in TeV emitting electrons in the nebula (see Eq. \ref{eq:recon_Lorentz}). However, several models with soft particle spectra ($\Gamma_{e} > 2.5$) and mG magnetic fields, show that the detection of flares at TeV energies can be achieved (Fig. \ref{fig:MAGIClikeflareintegralfluxes}). In fact, if the energy in electrons available is boosted above the upper limit established (due to re-acceleration of particles processes), the prospects for the detection of flares in the TeV regime with the CTA medium and small size telescopes (MSTs and SSTs, with best sensitivity at TeV energies) can significantly improve.

Observations with CTA will allow us to constrain many physical parameters in the acceleration and emission regions. We show that beside strong constraints on the particle spectrum and magnetic field, different hypothesis can be tested: A jitter mechanism could be easily proven if at work, whereas the dynamic of relativistic blobs can be derived if a MHD outflow is considered. For completeness, we discussed a third scenario, based on a non-MHD approach, that should be further theoretically investigated to infer more accurate predictions that can be compared with observations.

We conclude that even if the detection of the Crab Nebula flares in the TeV regime seems difficult to achieve, prospects to detect them in the sub-100 GeV regime are bright, and feasible in early stages of the CTA operation.

\section{Data Availability}
The data compilation for the Crab nebula was taken from \citealt{2010A&A...523A...2M} and \citealt{Buehler2012}. The CTA IRFs are public and available in \url{https://www.cta-observatory.org/science/cta-performance/} 


\section{Acknowledgements}

E. M. and D. F. T. acknowledge the support of the grants AYA2017-92402- EXP, PGC2018-095512-B-I00, iLink 2017-1238 (Ministerio de Econom\'ia y Competitividad) and SGR2017-1383 (Generalitat de Catalunya). We acknowledge the support of the
PHAROS COST Action (CA16214). E. de O. W. ackowledges the Alexander von Humboldt Foundation for finantial support. D. K. is supported by JSPS KAKENHI Grant Numbers JP18H03722, JP24105007, and JP16H02170.
We made use of the Cherenkov Telescope Array instrument response functions provided by the CTA Consortium and Observatory, see http://www.cta-observatory.org/science/cta-performance/ (version prod3b-v1) for more details. This paper has gone through an internal review by the CTA Consortium. This work was conducted in the context of the CTA Analysis and Simulation Working Group (ASWG).
%
%
We made use R Project for Statistical Computing (\citealt{Rmanual}).
Also, this research made use of \textsc{astropy}, a community-developed core \textsc{python} package for Astronomy (\textsc{astropy} Collaboration, 2018).

\bibliographystyle{mnras}

\bibliography{sample}

\begin{thebibliography}{}
\makeatletter
\relax
\def\mn@urlcharsother{\let\do\@makeother \do\$\do\&\do\#\do\^\do\_\do\%\do\~}
\def\mn@doi{\begingroup\mn@urlcharsother \@ifnextchar [ {\mn@doi@}
  {\mn@doi@[]}}
\def\mn@doi@[#1]#2{\def\@tempa{#1}\ifx\@tempa\@empty \href
  {http://dx.doi.org/#2} {doi:#2}\else \href {http://dx.doi.org/#2} {#1}\fi
  \endgroup}
\def\mn@eprint#1#2{\mn@eprint@#1:#2::\@nil}
\def\mn@eprint@arXiv#1{\href {http://arxiv.org/abs/#1} {{\tt arXiv:#1}}}
\def\mn@eprint@dblp#1{\href {http://dblp.uni-trier.de/rec/bibtex/#1.xml}
  {dblp:#1}}
\def\mn@eprint@#1:#2:#3:#4\@nil{\def\@tempa {#1}\def\@tempb {#2}\def\@tempc
  {#3}\ifx \@tempc \@empty \let \@tempc \@tempb \let \@tempb \@tempa \fi \ifx
  \@tempb \@empty \def\@tempb {arXiv}\fi \@ifundefined
  {mn@eprint@\@tempb}{\@tempb:\@tempc}{\expandafter \expandafter \csname
  mn@eprint@\@tempb\endcsname \expandafter{\@tempc}}}

\bibitem[\protect\citeauthoryear{{Abdalla} et~al.,}{{Abdalla}
  et~al.}{2019}]{2019Natur.575..464A}
{Abdalla} H.,  et~al., 2019, \mn@doi [\nat] {10.1038/s41586-019-1743-9}, \href
  {https://ui.adsabs.harvard.edu/abs/2019Natur.575..464A} {575, 464}

\bibitem[\protect\citeauthoryear{Abdo et~al.,}{Abdo et~al.}{2011}]{Abdo2011}
Abdo A.~A.,  et~al., 2011, \mn@doi [Science] {10.1126/science.1199705}, 331,
  739

\bibitem[\protect\citeauthoryear{{Abeysekara} et~al.,}{{Abeysekara}
  et~al.}{2019}]{2019Abeysekara}
{Abeysekara} A.~U.,  et~al., 2019, \mn@doi [\apj] {10.3847/1538-4357/ab2f7d},
  \href {https://ui.adsabs.harvard.edu/abs/2019ApJ...881..134A} {881, 134}

\bibitem[\protect\citeauthoryear{{Acharya} et~al.,}{{Acharya}
  et~al.}{2013}]{2013CTA}
{Acharya} B.~S.,  et~al., 2013, \mn@doi [Astroparticle Physics]
  {10.1016/j.astropartphys.2013.01.007}, \href
  {http://adsabs.harvard.edu/abs/2013APh....43....3A} {43, 3}

\bibitem[\protect\citeauthoryear{{Aharonian}, {Atoyan}  \&
  {Kifune}}{{Aharonian} et~al.}{1997}]{1997Aharonian}
{Aharonian} F.~A.,  {Atoyan} A.~M.,   {Kifune} T.,  1997, \mn@doi [\mnras]
  {10.1093/mnras/291.1.162}, \href
  {https://ui.adsabs.harvard.edu/abs/1997MNRAS.291..162A} {291, 162}

\bibitem[\protect\citeauthoryear{{Aharonian} et~al.,}{{Aharonian}
  et~al.}{2000}]{2000Aharonian}
{Aharonian} F.~A.,  et~al., 2000, \mn@doi [\apj] {10.1086/309225}, \href
  {https://ui.adsabs.harvard.edu/abs/2000ApJ...539..317A} {539, 317}

\bibitem[\protect\citeauthoryear{{Aharonian}, {Kelner}  \&
  {Prosekin}}{{Aharonian} et~al.}{2010}]{2010PhRvD..82d3002A}
{Aharonian} F.~A.,  {Kelner} S.~R.,   {Prosekin} A.~Y.,  2010, \mn@doi [\prd]
  {10.1103/PhysRevD.82.043002}, \href
  {https://ui.adsabs.harvard.edu/abs/2010PhRvD..82d3002A} {82, 043002}

\bibitem[\protect\citeauthoryear{{Aleksi{\'c}} et~al.,}{{Aleksi{\'c}}
  et~al.}{2015}]{Aleksic2015}
{Aleksi{\'c}} J.,  et~al., 2015, \mn@doi [Journal of High Energy Astrophysics]
  {10.1016/j.jheap.2015.01.002}, \href
  {http://adsabs.harvard.edu/abs/2015JHEAp...5...30A} {5, 30}

\bibitem[\protect\citeauthoryear{{Aleksi{\'c}} et~al.,}{{Aleksi{\'c}}
  et~al.}{2016}]{2016Aleksic}
{Aleksi{\'c}} J.,  et~al., 2016, \mn@doi [Astroparticle Physics]
  {10.1016/j.astropartphys.2015.02.005}, \href
  {https://ui.adsabs.harvard.edu/abs/2016APh....72...76A} {72, 76}

\bibitem[\protect\citeauthoryear{{Aliu} et~al.,}{{Aliu}
  et~al.}{2014}]{2014Aliu}
{Aliu} E.,  et~al., 2014, \mn@doi [\apjl] {10.1088/2041-8205/781/1/L11}, \href
  {https://ui.adsabs.harvard.edu/abs/2014ApJ...781L..11A} {781, L11}

\bibitem[\protect\citeauthoryear{Amenomori et~al.,}{Amenomori
  et~al.}{2019}]{2019Amenomori}
Amenomori M.,  et~al., 2019, \mn@doi [Phys. Rev. Lett.]
  {10.1103/PhysRevLett.123.051101}, 123, 051101

\bibitem[\protect\citeauthoryear{{Arakawa}, {Hayashida}, {Khangulyan}  \&
  {Uchiyama}}{{Arakawa} et~al.}{2020a}]{2020arXiv200507958A}
{Arakawa} M.,  {Hayashida} M.,  {Khangulyan} D.,   {Uchiyama} Y.,  2020a, arXiv
  e-prints, \href {https://ui.adsabs.harvard.edu/abs/2020arXiv200507958A} {p.
  arXiv:2005.07958}

\bibitem[\protect\citeauthoryear{{Arakawa}, {Hayashida}, {Khangulyan}  \&
  {Uchiyama}}{{Arakawa} et~al.}{2020b}]{2020Arakawa}
{Arakawa} M.,  {Hayashida} M.,  {Khangulyan} D.,   {Uchiyama} Y.,  2020b, arXiv
  e-prints, \href {https://ui.adsabs.harvard.edu/abs/2020arXiv200507958A} {p.
  arXiv:2005.07958}

\bibitem[\protect\citeauthoryear{Atoyan \& Aharonian}{Atoyan \&
  Aharonian}{1996}]{1996Atoyan}
Atoyan A.~M.,  Aharonian F.~A.,  1996, \mn@doi [\mnras]
  {10.1093/mnras/278.2.525}, 278, 525

\bibitem[\protect\citeauthoryear{Balbo, Walter, Ferrigno  \& Bordas}{Balbo
  et~al.}{2011}]{Balbo_2011}
Balbo M.,  Walter R.,  Ferrigno C.,   Bordas P.,  2011, \mn@doi [Astronomy &
  Astrophysics] {10.1051/0004-6361/201015980}, 527, L4

\bibitem[\protect\citeauthoryear{{Bartoli} et~al.,}{{Bartoli}
  et~al.}{2015}]{2015Bartoli}
{Bartoli} B.,  et~al., 2015, \mn@doi [\apj] {10.1088/0004-637X/798/2/119},
  \href {https://ui.adsabs.harvard.edu/abs/2015ApJ...798..119B} {798, 119}

\bibitem[\protect\citeauthoryear{{Bednarek} \& {Idec}}{{Bednarek} \&
  {Idec}}{2011}]{2011MNRAS.414.2229B}
{Bednarek} W.,  {Idec} W.,  2011, \mn@doi [\mnras]
  {10.1111/j.1365-2966.2011.18539.x}, \href
  {https://ui.adsabs.harvard.edu/abs/2011MNRAS.414.2229B} {414, 2229}

\bibitem[\protect\citeauthoryear{{Buehler}, {D'Ammando}  \& {Cannon}}{{Buehler}
  et~al.}{2011}]{2011ATelB}
{Buehler} R.,  {D'Ammando} F.,   {Cannon} A.,  2011, The Astronomer's Telegram,
  \href {https://ui.adsabs.harvard.edu/abs/2011ATel.3276....1B} {3276, 1}

\bibitem[\protect\citeauthoryear{Buehler et~al.,}{Buehler
  et~al.}{2012}]{Buehler2012}
Buehler R.,  et~al., 2012, \apj, 749, 26

\bibitem[\protect\citeauthoryear{{B{\"u}hler} \& {Blandford}}{{B{\"u}hler} \&
  {Blandford}}{2014}]{2014Buhler}
{B{\"u}hler} R.,  {Blandford} R.,  2014, \mn@doi [Reports on Progress in
  Physics] {10.1088/0034-4885/77/6/066901}, \href
  {https://ui.adsabs.harvard.edu/abs/2014RPPh...77f6901B} {77, 066901}

\bibitem[\protect\citeauthoryear{{Bykov}, {Pavlov}, {Artemyev}  \&
  {Uvarov}}{{Bykov} et~al.}{2012}]{2012Bykov}
{Bykov} A.~M.,  {Pavlov} G.~G.,  {Artemyev} A.~V.,   {Uvarov} Y.~A.,  2012,
  \mn@doi [\mnras] {10.1111/j.1745-3933.2011.01208.x}, \href
  {https://ui.adsabs.harvard.edu/abs/2012MNRAS.421L..67B} {421, L67}

\bibitem[\protect\citeauthoryear{{Cash}}{{Cash}}{1979}]{1979Cash}
{Cash} W.,  1979, \mn@doi [\apj] {10.1086/156922}, 228, 939

\bibitem[\protect\citeauthoryear{{Cerutti}, {Werner}, {Uzdensky}  \&
  {Begelman}}{{Cerutti} et~al.}{2012}]{2012ApJ...754L..33C}
{Cerutti} B.,  {Werner} G.~R.,  {Uzdensky} D.~A.,   {Begelman} M.~C.,  2012,
  \mn@doi [\apjl] {10.1088/2041-8205/754/2/L33}, \href
  {https://ui.adsabs.harvard.edu/abs/2012ApJ...754L..33C} {754, L33}

\bibitem[\protect\citeauthoryear{{Deil} et~al.}{{Deil}
  et~al.}{2017}]{2017gammapy}
{Deil} C.,  et~al., 2017, International Cosmic Ray Conference, 35, 766

\bibitem[\protect\citeauthoryear{{Derishev} \& {Aharonian}}{{Derishev} \&
  {Aharonian}}{2019}]{2019ApJ...887..181D}
{Derishev} E.,  {Aharonian} F.,  2019, \mn@doi [\apj]
  {10.3847/1538-4357/ab536a}, \href
  {https://ui.adsabs.harvard.edu/abs/2019ApJ...887..181D} {887, 181}

\bibitem[\protect\citeauthoryear{Fleishman}{Fleishman}{2006}]{Fleishman2006}
Fleishman G.~D.,  2006, \mn@doi [\apj] {10.1086/498732}, 638, 348

\bibitem[\protect\citeauthoryear{{Guilbert}, {Fabian}  \& {Rees}}{{Guilbert}
  et~al.}{1983}]{1983MNRAS.205..593G}
{Guilbert} P.~W.,  {Fabian} A.~C.,   {Rees} M.~J.,  1983, \mn@doi [Mon. Not. R.
  Astron. Soc.] {10.1093/mnras/205.3.593}, \href
  {https://ui.adsabs.harvard.edu/abs/1983MNRAS.205..593G} {205, 593}

\bibitem[\protect\citeauthoryear{{H.~E.~S.~S. Collaboration}
  et~al.,}{{H.~E.~S.~S. Collaboration} et~al.}{2014}]{2014HESS}
{H.~E.~S.~S. Collaboration} et~al., 2014, \mn@doi [\aap]
  {10.1051/0004-6361/201323013}, \href
  {https://ui.adsabs.harvard.edu/abs/2014A&A...562L...4H} {562, L4}

\bibitem[\protect\citeauthoryear{{Hassan} et~al.,}{{Hassan}
  et~al.}{2015}]{2015Hassan}
{Hassan} T.,  et~al., 2015, PoS ICRC2015

\bibitem[\protect\citeauthoryear{{Hillas} et~al.,}{{Hillas}
  et~al.}{1998}]{1998ApJ...503..744H}
{Hillas} A.~M.,  et~al., 1998, \mn@doi [\apj] {10.1086/306005}, \href
  {https://ui.adsabs.harvard.edu/abs/1998ApJ...503..744H} {503, 744}

\bibitem[\protect\citeauthoryear{{Holler} et~al.,}{{Holler}
  et~al.}{2015}]{2015Holler}
{Holler} M.,  et~al., 2015, PoS ICRC2015, \href
  {https://ui.adsabs.harvard.edu/abs/2015arXiv150902902H} {p. arXiv:1509.02902}

\bibitem[\protect\citeauthoryear{{Horns} \& {Aharonian}}{{Horns} \&
  {Aharonian}}{2004}]{2004Horns}
{Horns} D.,  {Aharonian} F.~A.,  2004, in {Schoenfelder} V.,  {Lichti} G.,
  {Winkler} C.,  eds,  ESA Special Publication Vol. 552, 5th INTEGRAL Workshop
  on the INTEGRAL Universe. p.~439 (\mn@eprint {arXiv} {astro-ph/0407119})

\bibitem[\protect\citeauthoryear{Kelner, Aharonian  \& Khangulyan}{Kelner
  et~al.}{2013}]{Kelner2013}
Kelner S.~R.,  Aharonian F.~A.,   Khangulyan D.,  2013, \mn@doi [\apj]
  {10.1088/0004-637x/774/1/61}, 774, 61

\bibitem[\protect\citeauthoryear{{Kennel} \& {Coroniti}}{{Kennel} \&
  {Coroniti}}{1984}]{1984ApJ...283..710K}
{Kennel} C.~F.,  {Coroniti} F.~V.,  1984, \mn@doi [\apj] {10.1086/162357},
  \href {http://adsabs.harvard.edu/abs/1984ApJ...283..710K} {283, 710}

\bibitem[\protect\citeauthoryear{{Khangulyan}, {Aharonian}  \&
  {Kelner}}{{Khangulyan} et~al.}{2014}]{2014ApJ...783..100K}
{Khangulyan} D.,  {Aharonian} F.~A.,   {Kelner} S.~R.,  2014, \mn@doi [\apj]
  {10.1088/0004-637X/783/2/100}, \href
  {https://ui.adsabs.harvard.edu/abs/2014ApJ...783..100K} {783, 100}

\bibitem[\protect\citeauthoryear{{Khangulyan}, {Aharonian}, {Romoli}  \&
  {Taylor}}{{Khangulyan} et~al.}{2020a}]{2020arXiv200300927K}
{Khangulyan} D.,  {Aharonian} F.,  {Romoli} C.,   {Taylor} A.,  2020a, arXiv
  e-prints, \href {https://ui.adsabs.harvard.edu/abs/2020arXiv200300927K} {p.
  arXiv:2003.00927}

\bibitem[\protect\citeauthoryear{{Khangulyan}, {Arakawa}  \&
  {Aharonian}}{{Khangulyan} et~al.}{2020b}]{2020MNRAS.491.3217K}
{Khangulyan} D.,  {Arakawa} M.,   {Aharonian} F.,  2020b, \mn@doi [\mnras]
  {10.1093/mnras/stz3261}, \href
  {https://ui.adsabs.harvard.edu/abs/2020MNRAS.491.3217K} {491, 3217}

\bibitem[\protect\citeauthoryear{{Kirk} \& {Giacinti}}{{Kirk} \&
  {Giacinti}}{2017}]{2017PhRvL.119u1101K}
{Kirk} J.~G.,  {Giacinti} G.,  2017, \mn@doi [\prl]
  {10.1103/PhysRevLett.119.211101}, \href
  {https://ui.adsabs.harvard.edu/abs/2017PhRvL.119u1101K} {119, 211101}

\bibitem[\protect\citeauthoryear{Lagarias, A.~Reeds, H.~Wright  \&
  Wright}{Lagarias et~al.}{1998}]{NEDMEAD1}
Lagarias J.,  A.~Reeds J.,  H.~Wright M.,   Wright P.,  1998, SIAM Journal on
  Optimization, 9, 112

\bibitem[\protect\citeauthoryear{Longair}{Longair}{1981}]{Longairbook}
Longair M.~S.,  1981, {High energy astrophysics: an informal introduction for
  students of physics and astrophysics}.
Cambridge Univ. Press, Cambridge, \url {https://cds.cern.ch/record/100003}

\bibitem[\protect\citeauthoryear{{Lyubarsky}}{{Lyubarsky}}{2012}]{2012MNRAS.427.1497L}
{Lyubarsky} Y.~E.,  2012, \mn@doi [\mnras] {10.1111/j.1365-2966.2012.22097.x},
  \href {https://ui.adsabs.harvard.edu/abs/2012MNRAS.427.1497L} {427, 1497}

\bibitem[\protect\citeauthoryear{{Lyutikov}, {Komissarov}, {Sironi}  \&
  {Porth}}{{Lyutikov} et~al.}{2018}]{2018JPlPh..84b6301L}
{Lyutikov} M.,  {Komissarov} S.,  {Sironi} L.,   {Porth} O.,  2018, \mn@doi
  [Journal of Plasma Physics] {10.1017/S0022377818000168}, \href
  {https://ui.adsabs.harvard.edu/abs/2018JPlPh..84b6301L} {84, 635840201}

\bibitem[\protect\citeauthoryear{{MAGIC Collaboration} et~al.,}{{MAGIC
  Collaboration} et~al.}{2019}]{2019Natur.575..455M}
{MAGIC Collaboration} et~al., 2019, \mn@doi [\nat] {10.1038/s41586-019-1750-x},
  \href {https://ui.adsabs.harvard.edu/abs/2019Natur.575..455M} {575, 455}

\bibitem[\protect\citeauthoryear{{Mart{\'\i}n}, {Torres}  \&
  {Rea}}{{Mart{\'\i}n} et~al.}{2012}]{2012Martin}
{Mart{\'\i}n} J.,  {Torres} D.~F.,   {Rea} N.,  2012, \mn@doi [\mnras]
  {10.1111/j.1365-2966.2012.22014.x}, \href
  {https://ui.adsabs.harvard.edu/abs/2012MNRAS.427..415M} {427, 415}

\bibitem[\protect\citeauthoryear{{Mayer}, {Buehler}, {Hays}, {Cheung}, {Dutka},
  {Grove}, {Kerr}  \& {Ojha}}{{Mayer} et~al.}{2013}]{2013Mayer}
{Mayer} M.,  {Buehler} R.,  {Hays} E.,  {Cheung} C.~C.,  {Dutka} M.~S.,
  {Grove} J.~E.,  {Kerr} M.,   {Ojha} R.,  2013, \mn@doi [\apjl]
  {10.1088/2041-8205/775/2/L37}, \href
  {https://ui.adsabs.harvard.edu/abs/2013ApJ...775L..37M} {775, L37}

\bibitem[\protect\citeauthoryear{{Meagher} \& {VERITAS
  Collaboration}}{{Meagher} \& {VERITAS Collaboration}}{2015}]{2015Meagher}
{Meagher} K.,  {VERITAS Collaboration} 2015, in 34th International Cosmic Ray
  Conference (ICRC2015). p.~792 (\mn@eprint {arXiv} {1508.06442})

\bibitem[\protect\citeauthoryear{{Mestre}, {de O{\~n}a Wilhelmi}, {Zanin},
  {Torres}  \& {Tibaldo}}{{Mestre} et~al.}{2020}]{2019Mestre}
{Mestre} E.,  {de O{\~n}a Wilhelmi} E.,  {Zanin} R.,  {Torres} D.~F.,
  {Tibaldo} L.,  2020, \mn@doi [\mnras] {10.1093/mnras/stz3421}, \href
  {https://ui.adsabs.harvard.edu/abs/2020MNRAS.492..708M} {492, 708}

\bibitem[\protect\citeauthoryear{{Meyer}, {Horns}  \& {Zechlin}}{{Meyer}
  et~al.}{2010}]{2010A&A...523A...2M}
{Meyer} M.,  {Horns} D.,   {Zechlin} H.-S.,  2010, \mn@doi [\aap]
  {10.1051/0004-6361/201014108}, \href
  {http://adsabs.harvard.edu/abs/2010A\%26A...523A...2M} {523, A2}

\bibitem[\protect\citeauthoryear{{Ojha}, {Hays}, {Buehler}  \& {Dutka}}{{Ojha}
  et~al.}{2013}]{2012Ojha}
{Ojha} R.,  {Hays} R.,  {Buehler} E.,   {Dutka} M.,  2013, The Astronomer's
  Telegram, \href {http://www.astronomerstelegram.org/?read=4855} {4855}

\bibitem[\protect\citeauthoryear{{Piron} et~al.}{{Piron}
  et~al.}{2001}]{2001Piron}
{Piron} F.,  et~al., 2001, \mn@doi [Astronomy and Astrophysics]
  {10.1051/0004-6361:20010798}, 374, 895

\bibitem[\protect\citeauthoryear{{Porth}, {Komissarov}  \& {Keppens}}{{Porth}
  et~al.}{2014}]{2014IJMPS..2860168P}
{Porth} O.,  {Komissarov} S.~S.,   {Keppens} R.,  2014, in International
  Journal of Modern Physics Conference Series. p. 1460168,
  \mn@doi{10.1142/S2010194514601689}

\bibitem[\protect\citeauthoryear{{R Core Team}}{{R Core Team}}{2013}]{Rmanual}
{R Core Team} 2013, R: A Language and Environment for Statistical Computing.
R Foundation for Statistical Computing, Vienna, Austria, \url
  {http://www.R-project.org/}

\bibitem[\protect\citeauthoryear{{Rudak} \& {Dyks}}{{Rudak} \&
  {Dyks}}{1998}]{1998Rudak}
{Rudak} B.,  {Dyks} J.,  1998, \mn@doi [\mnras]
  {10.1046/j.1365-8711.1998.01200.x}, \href
  {https://ui.adsabs.harvard.edu/abs/1998MNRAS.295..337R} {295, 337}

\bibitem[\protect\citeauthoryear{Striani et~al.,}{Striani
  et~al.}{2011}]{Striani2011}
Striani E.,  et~al., 2011, \mn@doi [\apj] {10.1088/2041-8205/741/1/l5}, 741, L5

\bibitem[\protect\citeauthoryear{{Striani} et~al.,}{{Striani}
  et~al.}{2013}]{2013Striani}
{Striani} E.,  et~al., 2013, \mn@doi [\apj] {10.1088/0004-637X/765/1/52}, \href
  {https://ui.adsabs.harvard.edu/abs/2013ApJ...765...52S} {765, 52}

\bibitem[\protect\citeauthoryear{Tavani et~al.,}{Tavani
  et~al.}{2011}]{Tavani2011}
Tavani M.,  et~al., 2011, \mn@doi [Science] {10.1126/science.1200083}, 331, 736

\bibitem[\protect\citeauthoryear{{Teraki} \& {Takahara}}{{Teraki} \&
  {Takahara}}{2013}]{2013ApJ...763..131T}
{Teraki} Y.,  {Takahara} F.,  2013, \mn@doi [\apj]
  {10.1088/0004-637X/763/2/131}, \href
  {https://ui.adsabs.harvard.edu/abs/2013ApJ...763..131T} {763, 131}

\bibitem[\protect\citeauthoryear{Weisskopf et~al.,}{Weisskopf
  et~al.}{2013}]{Weisskopf2013}
Weisskopf M.~C.,  et~al., 2013, \apj, 765, 56

\bibitem[\protect\citeauthoryear{Wright}{Wright}{1996}]{NEADMED2}
Wright M.,  1996, Direct search methods: Once scorned, now respectable.
Addison-Wesley, pp 191--208

\bibitem[\protect\citeauthoryear{{Zabalza}}{{Zabalza}}{2015}]{naima}
{Zabalza} V.,  2015, Proc.~of International Cosmic Ray Conference 2015, \href
  {http://adsabs.harvard.edu/abs/2015arXiv150903319Z} {p.~922}

\bibitem[\protect\citeauthoryear{{Zenitani}, {Hesse}  \& {Klimas}}{{Zenitani}
  et~al.}{2009}]{2009ApJ...696.1385Z}
{Zenitani} S.,  {Hesse} M.,   {Klimas} A.,  2009, \mn@doi [\apj]
  {10.1088/0004-637X/696/2/1385}, \href
  {https://ui.adsabs.harvard.edu/abs/2009ApJ...696.1385Z} {696, 1385}

\makeatother
\end{thebibliography}

\end{document}